\documentclass[11pt,a4paper, oneside]{article}

\usepackage{amsmath}
\usepackage{amsfonts}
\usepackage{amssymb}
\usepackage{amscd}
\usepackage{latexsym}
\usepackage{mathrsfs}
\usepackage{amsthm}

\newcommand{\CC}{\mathbb{C}}
\newcommand{\RR}{\mathbb{R}}
\newcommand{\ZZ}{\mathbb{Z}}

\newcommand{\SSS}{\mathbb{S}}
\newcommand{\HH}{\mathcal{H}}
\newcommand{\dom}{\mathop{\mathcal{D}}}

\newcommand{\tr}{\mathop{\mathrm{tr}}}

\newcommand{\R}{{\mathbb R}}
\newcommand{\T}{{\mathbb T}}
\newcommand{\Z}{{\mathbb Z}}

\newcommand{\C}{{\mathbb C}}

\newcommand{\Fc}{{\mathcal F}}

\def\e{\varepsilon}
\def\L{\mathcal L}
\def\f{\mathfrak{f}}
\def\g{\mathfrak{g}}

\def\k{\textsc k}

\def\V{\mathcal V}

\def\d{\mathrm{d}}
\def\det{\mathrm{det}}
\def\AB{A\!B}
\def\CD{C\!D}
\def\F21{{}_2F_1}
\def\2{\mathfrak{int}}

\def\li{\mathrm{l.i.m.}}
\def\P{{\mathrm P}}
\def\PP{\mathcal{P}}

\newtheorem{theorem}{Theorem}
\newtheorem{prop}[theorem]{Proposition}
\newtheorem{lemma}[theorem]{Lemma}

\theoremstyle{definition}

\newtheorem{rem}[theorem]{\bf Remark}

\begin{document}

\title{\bf Spectral and scattering theory\\for the Aharonov-Bohm operators}

\author{Konstantin Pankrashkin$\,^1$ and Serge Richard$\,^2$}
\date{\small}
\maketitle

\begin{quote}
\begin{itemize}
\item[$^1$] Laboratoire de Math\'ematiques d'Orsay, CNRS UMR 8628, Universit\'e Paris-Sud XI, B\^atiment 425, 91405 Orsay Cedex, France;\\
E-mail: {\tt konstantin.pankrashkin@math.u-psud.fr}
\item[$^2$] Department of Pure Mathematics and Mathematical Statistics,
Centre for Mathematical Sciences, University of Cambridge,
Cambridge, CB3 0WB, United Kingdom; E-mail: {\tt sr510@cam.ac.uk}
\\[\smallskipamount]
On leave from Universit\'e de Lyon, Universit\'e Lyon I, CNRS UMR5208, Institut Camille Jordan, 43 blvd du 11 novembre 1918, 69622 Villeurbanne Cedex, France
\end{itemize}
\end{quote}

\begin{abstract}
\noindent We review the spectral and the scattering theory for the Aharonov-Bohm model on
$\RR^2$. New formulae for the wave operators and for the scattering operator are
presented. The asymptotics at high and at low energy of the scattering operator are computed.
\end{abstract}

\section{Introduction}

The Aharonov-Bohm (A-B) model
describing the motion of a charged particle
in a magnetic field concentrated at a single point
is one of the few systems in mathematical physics for which
the spectral and the scattering properties can be completely computed. It has been introduced
in \cite{AB} and the first rigorous treatment appeared in \cite{Rui}. A more general class
of models involving boundary conditions at the singularity point
has then been developed in \cite{AT,DS} and further extensions or refinements
appeared since these simultaneous works.
Being unable to list all these subsequent papers, let us simply mention few of them~:
\cite{T2} in which it is proved that
the A-B models can be obtained as limits in a suitable sense of systems with less singular
magnetic fields, and \cite{T1} in which it is shown that the low energy behavior of the
scattering amplitude for two dimensional magnetic Schr\"odinger operators is similar to
the scattering amplitude of the A-B models.
Concerning the extensions we mention the papers \cite{ESV} which considers the A-B operators
with an additional uniform magnetic field and \cite{OL} which studies the A-B operators
on the hyperbolic plane.

The aim of the present paper is to provide the spectral and the scattering analysis
of the A-B operators on $\RR^2$ for all possible values of the parameters (boundary conditions).
The work is motivated by the recent result of one of the authors \cite{R}
showing that the A-B wave operators
can be rewritten in terms of explicit functions of the generator of dilations and of the
Laplacian.
However, this astonishing result was partially obscured by some too complicated
expressions for the scattering operator borrowed from \cite{AT} and by a certain function
presented only in terms of its Fourier transform.
 For those reasons, we have decided to start again the analysis from scratch
 using the modern operator-theoretical machinery.
For example, our computations do not involve an explicit parametrization of $U(2)$ which
leads in \cite{AT} or in \cite{DS} to some unnecessary complications.
Simultaneously, we recast this analysis in the up-to-date theory of self-adjoint extensions \cite{BGP}
and derive rigorously the expressions for the wave operators and the scattering operator
from the stationary approach of scattering theory as presented in \cite{Y}.

So let us now describe the content of this review paper.
In Section \ref{secgeneral} we introduce the operator $H_\alpha$
which corresponds to a Schr\"odinger operator in $\RR^2$ with a $\delta$-type
magnetic field at the origin. The index $\alpha$ corresponds to the total flux of the
magnetic field, and on a natural domain this operator has deficiency indices $(2,2)$.
The description of this natural domain is recalled and some of its properties are exhibited.

Section \ref{boundaryt} is devoted to the description of all self-adjoint extensions of the
operator $H_\alpha$.
More precisely, a boundary triple for the operator $H_\alpha$ is constructed in Proposition
\ref{bt}. It essentially consists in the definition of two linear maps $\Gamma_1,\Gamma_2$
from the domain $\dom(H_\alpha^*)$ of the adjoint of $H_\alpha$ to $\CC^2$ which have
some specific properties with respect to $H_\alpha$, as recalled at the beginning of this section.
Once these maps are exhibited, all self-adjoint extensions of $H_\alpha$ can be labeled by
two $2\times 2$-matrices $C$ and $D$ satisfying two simple conditions presented in \eqref{eq-mcd}.
These self-adjoint extensions are denoted by $H_\alpha^{\CD}$.
The $\gamma$-field and the Weyl function corresponding to the boundary triple are then constructed.
By taking advantage of some general results related to the boundary triple's approach,
they allow us to explicit the spectral properties of $H_\alpha^{\CD}$ in very simple terms.
At the end of the section we add some comments about the role of the parameters $C$ and $D$
and discuss some of their properties.

The short Section \ref{secFour} contains formulae on the Fourier transform and on the
dilation group that are going to be used subsequently.
Section \ref{secSca} is the main section on scattering theory.
It contains the time dependent approach as well as the stationary approach of the
scattering theory for the A-B models. Some calculations involving Bessel functions
or hypergeometric $\F21$-functions look rather tricky but they are necessary for
a rigorous derivation of the stationary expressions. Fortunately,
the final expressions are much more easily understandable.
For example, it is proved in Proposition \ref{waveopAB} that the channel wave operators
for the original A-B operator $H_\alpha^{\AB}$ are equal to very explicit functions
of the generator of dilation. These functions are continuous on $[-\infty,\infty]$ and take
values in the set of complex number of modulus $1$.
Theorem \ref{formulewave} contains a similar explicit description of the wave operators
for the general operator $H_\alpha^{\CD}$.

In Section \ref{secScatOp} we study the scattering operator and in particular its
asymptotics at small and large energies. These properties highly depend
on the parameters $C$ and $D$ but also on the flux $\alpha$ of the singular magnetic field.
All the various possibilities are explicitly analysed. The statement looks rather messy,
but this simply reflects the richness of the model.

The parametrization of the self-adjoint extensions of $H_\alpha$ with the pair $(C,D)$ is highly
non unique. For convenience, we introduce in the last section a one-to-one parametrization
of all self-adjoint extensions and explicit some of the previous results in this framework.
For further investigations in the structure of the set of all self-adjoint extensions,
this unique parametrization has many advantages.

Finally, let us mention that this paper is essentially self-contained.
Furthermore, despite the rather long and rich history of the Aharonov-Bohm
model most of the our results are new or exhibited in the present form for the first time.

\begin{rem}
After the completion of this paper, the authors were informed about the closely related work \cite{BDG}. In this paper, the differential expression
$-\partial_x^2 + (m^2-1/4)x^{-2}$ on $\RR_+$ is considered and a holomorphic family
of extensions for $\Re(m)>-1$ is studied. Formulae for the wave operators
similar to our formula \eqref{niceGamma} were independently
obtained by its authors.
\end{rem}

\section{General setting}\label{secgeneral}

Let $\HH$ denote the Hilbert space $L^2(\RR^2)$ with its scalar product $\langle \cdot,\cdot\rangle$ and its norm $\|\cdot \|$. For any $\alpha \in \RR $, we set $A_\alpha: \RR^2\setminus\{0\} \to \RR^2$ by
\begin{equation*}
A_\alpha(x,y)= -\alpha \Big(\frac{-y}{x^2+y^2},
\frac{x}{x^2+y^2}
\Big),
\end{equation*}
corresponding formally to the magnetic field $B=\alpha\delta$ ($\delta$ is the Dirac delta function),
and consider the operator
\begin{equation*}
H_\alpha:=(-i\nabla -A_\alpha)^2,
\qquad \dom(H_\alpha)=C_c^\infty\big(\RR^2\setminus\{0\}\big)\ .
\end{equation*}
Here $C_c^\infty(\Xi)$ denotes the set of smooth functions on $\Xi$ with compact support.
The closure of this operator in $\HH$, which is denoted by the same symbol, is symmetric and has deficiency indices $(2,2)$ \cite{AT,DS}.
For further investigation we need some more information
on this closure.

So let us first decompose the Hilbert space $\HH$ with respect to polar coordinates: For any $m \in \ZZ$, let $\phi_m$ be the complex function defined by
$[0,2\pi)\ni \theta \mapsto \phi_m(\theta):= \dfrac{e^{im\theta}}{\sqrt{2\pi}}$. Then, by taking the completeness of the family $\{\phi_m\}_{m \in \ZZ}$ in $L^2(\SSS^1)$ into account, one has the canonical isomorphism
\begin{equation}\label{decomposition}
\HH \cong \bigoplus_{m \in \ZZ} \HH_r \otimes [\phi_m] \ ,
\end{equation}
where $\HH_r:=L^2(\RR_+, r\;\!\d r)$ and $[\phi_m]$ denotes the one dimensional space spanned by $\phi_m$. For shortness, we write $\HH_m$ for $\HH_r \otimes [\phi_m]$, and often consider it as a subspace of $\HH$. Clearly, the Hilbert space $\HH_m$ is isomorphic to $\HH_r$, for any $m$

In this representation the operator $H_\alpha$ is equal to \cite[Sec.~2]{DS}
\begin{equation} \label{dec-S}
\bigoplus_{m\in\ZZ} H_{\alpha,m} \otimes 1,
\end{equation}
with
\begin{equation*}
H_{\alpha,m} = - \frac{\d^2}{\d r^2}-\frac{1}{r}\,\frac{\d}{\d r}+\frac{(m+\alpha^2)}{r^2},
\end{equation*}
and with a domain which depends on $m+\alpha$.
It clearly follows from this representation that replacing $\alpha$ by $\alpha+n$, $n\in\ZZ$,
corresponds to a unitary transformation of $H_\alpha$. In particular,
the case $\alpha\in\ZZ$ is equivalent to the magnetic field-free case $\alpha=0$,
{\it i.e.}~the Laplacian and its zero-range perturbations, see \cite[Chapt.~1.5]{AGHH}.
Hence throughout the paper we restrict our attention to the values $\alpha\in(0,1)$.

So, for $\alpha\in(0,1)$ and $m \not \in \{0,-1\}$, the domain $\dom(H_{\alpha,m})$ is given by
\begin{equation*}
\Big\{f\in \HH_r\cap \HH^{2,2}_{loc}(\RR_+)\mid
-f''-r^{-1}f'+(m+\alpha)^2r^{-2} f\in \HH_r \Big\} \ .
\end{equation*}
For $m\in\{0,-1\}$, let $H_\nu^{(1)}$ denote
the Hankel function of the first kind and of order $\nu$, and for $f,h \in \HH^{2,2}_{loc}$ let $W(g,h)$ stand for the Wronskian
\begin{equation*}
W(f,h):= \overline{f} h' - \overline{f'} h \ .
\end{equation*}
One then has
\begin{multline*}
\dom(H_{\alpha,m})=
\Big\{f\in \HH_r\cap \HH^{2,2}_{loc}(\RR_+)\mid\,\\
-f''-r^{-1}f'+(m+\alpha)^2r^{-2} f\in \HH_r \hbox{ and }
\lim_{r\to 0+}r\big[W(f,h_{\pm i,m})\big](r)=0
\Big\},
\end{multline*}
where
$h_{+i,m}(r)=H^{(1)}_{|m+\alpha|} (e^{i\pi/4} r)$ and
$h_{-i,m}(r)=H^{(1)}_{|m+\alpha|} (e^{i3\pi/4} r)$.
It is known that the operator $H_{\alpha,m}$ for $m \notin \{0,-1\}$ are self-adjoint on the mentioned domain, while $H_{\alpha,0}$ and $H_{\alpha,-1}$ have deficiency indices $(1,1)$.
This explains the deficiency indices $(2,2)$ for the operator $H_\alpha$.

The problem of the description of all self-adjoint extensions of the operator $H_\alpha$ can be approached by two methods. On the one hand, there exists the classical description of von Neumann based on unitary operators between the deficiency subspaces. On the other hand, there exists the theory of boundary triples which has been widely developed for the last twenty years \cite{BGP,DM}. Since our construction is based only on the latter approach, we shall recall it briefly in the sequel.

Before stating a simple result on $\dom(H_{\alpha,m})$ for $m \in \{0,-1\}$ let us set some conventions. For a complex number $z\in\CC\setminus\RR_+$, the branch of the square root $z\mapsto\sqrt{z}$
is fixed by the condition $\Im \sqrt z>0$. In other words, for $z=r e^{i\varphi}$ with $r>0$
and $\varphi\in(0,2\pi)$ one has $\sqrt z=\sqrt r e^{i\varphi/2}$. On the other hand, for $\beta\in \RR$
we always take the principal branch of the power $z\mapsto\to z^\beta$
by taking the principal branch of the argument $\arg z\in (-\pi,\pi)$.
This means that for $z=r e^{i\varphi}$ with $r>0$
and $\varphi\in(-\pi,\pi)$ we have $z^\beta=r^\beta e^{i\beta\varphi}$.
Let us also recall the asymptotic behavior of $H^{(1)}_\nu(w)$ as $w \to 0$ in $\C\setminus \RR_-$ and for $\nu \not \in \ZZ$:
\begin{equation}
 \label{eq-asymp}
H^{(1)}_\nu(w)=-\frac{2^\nu i}{\sin(\pi\nu) \;\!\Gamma(1-\nu)}\, w^{-\nu}
+\frac{2^{-\nu} i e^{-i\pi\nu}}{\sin (\pi \nu)\;\! \Gamma(1+\nu)} \, w^\nu
+O(w^{2-\nu}).
\end{equation}

\begin{prop}\label{prop1}
For any $f\in \dom(H_{\alpha,m})$ with $m\in\{0,-1\}$, the following asymptotic behavior holds:
\begin{equation*}
\lim_{r\to 0+} \frac{f(r)}{r^{|m+\alpha|}}=0.
\end{equation*}

\begin{proof}
Let us set $\nu:=|m+\alpha|\in (0,1)$, and recall that $f\in\dom(H_{\alpha,m})$ implies $f\in C^1\big((0,+\infty)\big)$ and that
the Hankel function satisfies
$\big(H^{(1)}_\nu(z)\big)'=H^{(1)}_{\nu-1}(z)-\frac{\nu}{z}H^{(1)}_\nu(z)$.
By taking this and the asymptotics \eqref{eq-asymp} into account,
the condition $\lim_{r\to 0+} r[W(h_{\pm i,m},f)](r)=0$
implies that
\begin{equation}\label{eq-n1}
\lim_{r\to 0+} \big\{r^{\nu+1} f'(r)-\nu r^{\nu} f(r)\big\}=0
\end{equation}
and that
\begin{equation}\label{eq-n2}
\lim_{r\to 0+}\big\{r^{1-\nu} f'(r)+\nu r^{-\nu} f(r)\big\}=0\ .
\end{equation}
Multiplying both terms of \eqref{eq-n2} by $r^{2\nu}$ and subtracting it
from \eqref{eq-n1} one obtains that
\begin{equation}
  \label{eq-as2}
\lim_{r\to 0+} r^{\nu} f(r)=0.
\end{equation}
On the other hand, considering \eqref{eq-n2} as a linear differential equation for $f$:
$r^{1-\nu} f'(r)+\nu r^{-\nu} f(r)=b(r)$,
and using the variation of constant one gets for some $C \in \C$:
\[
f(r)= \frac{C}{r^\nu} +\frac{1}{r^\nu}\,\int_0^r t^{2\nu-1}\;\!b(t)\;\!\d t\ .
\]
Now Eq.~\eqref{eq-as2} implies that $C=0$, and by using l'H\^opital's rule, one finally obtains:
\[
\lim_{r\to 0+}\frac{f(r)}{r^\nu}=
\lim_{r\to 0+}\frac{\displaystyle \int_0^r t^{2\nu -1}\;\!b(t)\;\!\d t}{r^{2\nu}}
=\lim_{r\to 0+}\frac{r^{2\nu -1}b(r)}{2\,\nu \,r^{2\nu-1}}
=\frac{1}{2\,\nu}\,\lim_{r\to 0+}b(r)=0.
\qedhere
\]
\end{proof}
\end{prop}

\section{Boundary conditions and spectral theory}\label{boundaryt}

In this section, we explicitly construct a boundary triple for the operator $H_\alpha$ and we briefly exhibit some spectral results in that setting. Clearly, our construction is very closed to the one in \cite{DS}, but this paper does not contain any reference to the boundary triple
machinery. Our aim is thus to recast the construction in an up-to-date theory.
The following presentation is strictly adapted to our setting, and as a general rule we omit to write the dependence on $\alpha$ on each of the objects. We refer to \cite{BGP} for more information on boundary triples.

Let $H_\alpha$ be the densely defined closed and symmetric operator in $\HH$ previously introduced. The adjoint of $H_\alpha$ is denoted by $H_\alpha^*$ and is defined on the domain
\begin{equation*}
\dom(H_\alpha^*) = \Big\{
\f \in \HH \cap H^{2,2}_{loc}\big(\RR^2\setminus\{0\}\big) \mid H_\alpha \f \in \HH \Big\}.
\end{equation*}
Let $\Gamma_1$, $\Gamma_2$ be two linear maps from $\dom(H_\alpha^*)$ to $\CC^2$. The triple $(\CC^2,\Gamma_1,\Gamma_2)$ is called a \emph{boundary triple for $H_\alpha$} if the following two conditions are satisfied:
\begin{enumerate}
\item[(1)] $\langle  \f, H_\alpha^* \;\!\g\rangle - \langle H_\alpha^* \;\!\f, \g \rangle = \langle \Gamma_1 \;\!\f,\Gamma_2\;\! \g\rangle-\langle \Gamma_2\;\! \f,\Gamma_1\;\!\g \rangle$ for any $\f,\g \in \dom(H_\alpha^*)$,
\item[(2)] the map $(\Gamma_1,\Gamma_2):\dom(H_\alpha^*)\to \CC^2 \oplus \CC^2$ is surjective.
\end{enumerate}

It is proved in  the reference mentioned above that such a boundary triple exists, and that all self-adjoint extensions of $H_\alpha$ can be described in this framework. More precisely, let $C, D\in M_2(\CC)$ be $2\times 2$ matrices, and let us denote by $H_\alpha^{\CD}$  the restriction of $H_\alpha^*$ on the domain
\begin{equation*}
\dom(H_\alpha^{\CD}):=\{\f \in \dom(H^*_\alpha)\mid C\Gamma_1 \f = D \Gamma_2 \f\}\ .
\end{equation*}
Then, the operator $H_\alpha^{\CD}$ is self-adjoint if and only if the matrices $C$ and $D$ satisfy the following conditions:
\begin{equation}
\label{eq-mcd}
\text{(i) $CD^*$ is self-adjoint,\qquad  (ii) $\det(CC^* + DD^*)\neq 0$.}
\end{equation}
Moreover, any self-adjoint extension of $H_\alpha$ in $\HH$ is equal to one of the operator $H_\alpha^{\CD}$.

We shall now construct explicitly a boundary triple for the operator $H_\alpha$. For that purpose, let us consider $z\in \CC \setminus \RR_+$ and choose $k =\sqrt{z}$ with $\Im(k)>0$. It is easily proved that the following two functions $\f_{z,0}$ and $\f_{z,-1}$ define an orthonormal  basis in $\ker (H_\alpha^*-z)$, namely in polar coordinates:
\begin{equation*}
\f_{z,0}(r,\theta)=N_{z,0}\;\!H^{(1)}_\alpha (kr) \;\!\phi_0(\theta),
\quad
\f_{z,-1}(r,\theta)=N_{z,-1}\;\!H^{(1)}_{1-\alpha} (k r) \;\!\phi_{-1}(\theta),
\end{equation*}
where $N_{z,m}$ is the normalization such that $\|f_{z,0}\|=\|f_{z,-1}\|=1$.
In particular, by making use of the equality
\begin{equation*}
\int_0^\infty r \big|H^{(1)}_\nu(kr)\big|^2 \d r=\big(\pi\cos(\pi \nu/2)\big)^{-1}
\end{equation*}
valid for $k\in \{e^{i\pi/4},e^{i3\pi/4}\}$, one has
\begin{equation*}
N_{\pm i,0} = \big(\pi\cos(\pi \alpha/2)\big)^{1/2} \quad\hbox{ and }\quad
N_{\pm i,-1} = \big(\pi\cos(\pi (1-\alpha)/2)\big)^{1/2} = \big(\pi\sin(\pi \alpha/2)\big)^{1/2}\ .
\end{equation*}

Let us also introduce the averaging operator $\PP$ with respect to the polar angle acting on any $\f\in \HH$ and for almost every $r>0$ by
\begin{equation*}
[\PP(\f)](r)=\int_0^{2\pi} \f(r,\theta)\, \d\theta.
\end{equation*}
Following \cite[Sec.~3]{DS} we can then define the following four linear functionals on suitable $\f$:
\begin{align*}
\Phi_0 (\f)&=\lim_{r\to 0+} r^\alpha [\PP(\f\overline{\phi_0})](r), &
\Psi_0 (\f)&=\lim_{r\to 0+} r^{-\alpha} \big( [\PP(\f\overline{\phi_0})](r) - r^{-\alpha} \Phi_0 (\f)\big),\\
\Phi_{-1} (\f)&=\lim_{r\to 0+} r^{1-\alpha} [\PP( \f \overline{\phi_{-1}})] (r),&
\Psi_{-1} (\f)&=\lim_{r\to 0+} r^{\alpha-1} \big( [\PP(\f \overline{\phi_{-1}})](r) - r^{\alpha-1} \Phi_{-1}(\f)\big).
\end{align*}
For example, by taking the asymptotic behavior \eqref{eq-asymp} into account one obtains
\begin{equation}
   \label{eq-phipsi}
\begin{aligned}
\Phi_0 (\f_{z,0})&=N_{z,0}\;a_\alpha(z), & \Phi_{-1} (\f_{z,0})& =0, \\
\Psi_0 (\f_{z,0})&=N_{z,0}\;b_\alpha(z), &
\Psi_{-1} (\f_{z,0})&=0,\\
\Phi_{-1} (\f_{z,-1})&= N_{z,-1}\;a_{1-\alpha}(z), &
\Phi_0 (\f_{z,-1})&=0, \\
\Psi_{-1} (\f_{z,-1})&=N_{z,-1}\;b_{1-\alpha}(z), &
\Psi_0 (\f_{z,-1})&=0,
\end{aligned}
\end{equation}
with
\begin{equation}\label{horreur}
a_\nu(z)= -\frac{2^\nu i}{\sin (\pi\nu)\;\! \Gamma(1-\nu)}\;\!k^{-\nu},
\quad
b_\nu(z)= \frac{2^{-\nu} i e^{-i\pi\nu}}{\sin (\pi \nu)\;\! \Gamma(1+\nu)}\;\! k^\nu.
\end{equation}

The main result of this section is:

\begin{prop}\label{bt}
The triple $(\CC^2,\Gamma_1,\Gamma_2)$, with $\Gamma_1, \Gamma_2$ defined on $\f\in \dom(H_\alpha^*)$ by
\begin{equation*}
\Gamma_1 \f:=\left(\begin{smallmatrix}
\Phi_0 (\f)\\ \Phi_{-1} (\f)
\end{smallmatrix}\right),
\quad
\Gamma_2 \f:=
2
\left(\begin{smallmatrix}
\alpha\, \Psi_0 (\f)\\ (1-\alpha)\,\Psi_{-1}(\f)
\end{smallmatrix}\right),
\end{equation*}
is a boundary triple for $H_\alpha$.
\end{prop}

\begin{proof} We use the schema from \cite[Lem.~5]{BG}.
For any $\f,\g \in \dom(H_\alpha^*)$ let us define the sesquilinear forms
\begin{equation*}
B_1(\f,\g):=\langle \f, H_\alpha^*\g \rangle -\langle H_\alpha^*\f, \g \rangle
\end{equation*}
and
\begin{equation*}
B_2(\f,\g):=\langle \Gamma_1 \f\;\!,\Gamma_2 \;\! \g\rangle\ - \langle \Gamma_2\;\! \f,\Gamma_1 \;\! \g\rangle .
\end{equation*}
We are going to show that these expressions are well defined and that $B_1=B_2$.

i) Clearly, $B_1$ is well defined. For $B_2$, let us first recall that $\dom (H_\alpha^*)=\dom(H_\alpha) +\ker(H_\alpha^*-i)+\ker(H_\alpha^*+i)$.
It has already been proved above that the four maps $\Phi_{0}, \Phi_{-1}, \Psi_0$ and $\Psi_{-1}$ are well defined on the elements of $\ker(H_\alpha^*-i)$ and $\ker(H_\alpha^*+i)$.
We shall now prove that $\Gamma_1 \f=\Gamma_2 \f=0$ for $\f \in \dom(H_\alpha)$, which shows that $B_2$ is also well defined on $\dom(H_\alpha^*)$.
In view of the decomposition \eqref{dec-S} it is sufficient to consider functions $\f$ of the form $\f(r,\theta)=f_m(r) \phi_m(\theta)$ for any $m\in\ZZ$ and with $f_m\in\dom (H_{\alpha,m})$.
Obviously, for such a function $\f$ with $m\notin\{0,-1\}$ one has $[P(\f)](r)=0$ for almost every $r$, and thus $\Gamma_1 \f=\Gamma_2 \f=0$. For $m\in\{0,1\}$ the equalities
$\Gamma_1 \f=\Gamma_2 \f=0$ follow directly from
Proposition \ref{prop1}.

ii) Now, since $\Gamma_1 \f=\Gamma_2 \g=0$ for all $\f,\g\in\dom (H_\alpha)$, the only non trivial contributions to the sesquilinear form $B_2$ come from $\f,\g \in \ker(H_\alpha^*-i) + \ker(H_\alpha^*+i)$.
On the other hand one also has $B_1(\f,\g)=0$ for $\f,\g \in \dom(H_\alpha)$.
Thus, we are reduced in proving the equalities
\begin{equation*}
B_1(\f_{z, m},\f_{z', n})=B_2(\f_{z,m},\f_{z', n})
\end{equation*}
for any $z,z'\in\{-i,i\}$ and $m,n\in\{0,-1\}$.

Observe first that for $z \neq z'$ and arbitrary $m,n$ one has
\begin{equation*}
B_1(\f_{z,m},\f_{z',n})=
\langle \f_{z,m}, z' \f_{z',n}\rangle -
\langle z\f_{z,m},  \f_{z',n} \rangle=0
\end{equation*}
since $z'=\overline{z}$.
Now, for $m\ne n$
one has $\Gamma_1 \f_{z,m} \perp \Gamma_2 \f_{z',n}$, and
hence $B_2(\f_{z,m},\f_{z',n})=0=B_1(\f_{z,m},\f_{z',n})$.
For $m=n$ one easily calculate with $\nu:=|m-\alpha|$ that
\begin{equation*}
B_2(\f_{z,m},\f_{z',m})= 2 \nu \overline{N_{z,m}}\;\! N_{z',m}\big(
\overline{a_\nu(z)}\;\! b_\nu(z')-
\overline{b_\nu(z)}\;\!a_\nu(z')
\big)=0 \ ,
\end{equation*}
and then $B_2(\f_{z,m},\f_{z',m})=0=B_1(\f_{z,m},\f_{z',m})$.

We now consider $z=z'$ and $m\ne n$. One has
\begin{equation*}
B_1(\f_{z,m},\f_{z,n})=
\langle \f_{z,m}, z \f_{z,n}\rangle -
\langle z\f_{z,m},  \f_{z,n} \rangle=2z
\langle \f_{z,m}, \f_{z,n}\rangle=0
\end{equation*}
and again $\Gamma_1 \f_{z,m} \perp \Gamma_2 \f_{z,n}$. It then follows that $B_2(\f_{z,m},\f_{z,n})=0=B_1(\f_{z,m},\f_{z,n})$.

So it only remains to show that
$B_1(\f_{z,m},\f_{z,m})=B_2(\f_{z,m},\f_{z,m})$.
For that purpose, observe first that
\begin{equation*}
B_1(\f_{z,m},\f_{z,m}) = 2z \langle \f_{z,m}, \f_{z,m}\rangle=2z.
\end{equation*}
On the other hand, one has
\begin{equation*}
B_2( \f_{z,m},\f_{z,m}) =
2i \Im\big(\langle \Gamma_1 \f_{z,m},\Gamma_2 \f_{z,m}\big)
= 2i \Im\Big(
2 \nu |N_{z,m}|^2\;\!\overline{a_\nu(z)}\;\!b_\nu(z) \Big)
\end{equation*}
with $\nu = |m-\alpha|$.
By inserting \eqref{horreur} into this expression, one obtains (with $k =\sqrt{z}$ and $\Im(k)>0$)
\begin{eqnarray*}
B_2( \f_{z,m},\f_{z,m})& = &4i \nu |N_{z,m}|^2 \Im\Big(
\frac{-(k^\nu)^2\;\!e^{-i\pi\nu}}{\sin^2(\pi\nu)\;\!\Gamma(1-\nu)\;\! \Gamma(1+\nu)} \Big) \\
& = & 4z \nu |N_{z,m}|^2
\frac{\sin(\pi\nu/2)}{\sin^2(\pi\nu)\;\!\Gamma(1-\nu)\;\! \Gamma(1+\nu)}\ .
\end{eqnarray*}
Finally, by taking the equality
\begin{equation*}
\Gamma(1-\nu)\Gamma(1+\nu)=\frac{\pi \nu}{\sin (\pi\nu)}
\end{equation*}
into account, one obtains
\begin{equation*}
B_2( \f_{z,m},\f_{z,m}) = 4z |N_{z,m}|^2 \frac{\sin(\pi\nu/2)}{\sin(\pi\nu)\;\!\pi}
= 4z \pi\;\!\cos(\pi \nu/2)\;\!
\frac{\sin(\pi\nu/2)}{\sin(\pi\nu)\;\!\pi} =2z\ ,
\end{equation*}
which implies $B_2( \f_{z,m},\f_{z,m}) = 2z = B_1(\f_{z,m},\f_{z,m})$.

iii) The surjectivity of the map $(\Gamma_1,\Gamma_2):\dom(H^*_\alpha)\to \CC^2 \oplus \CC^2$
follows from the equalities \eqref{eq-phipsi}.
\end{proof}

Let us now construct \emph{the Weyl function} corresponding to the above boundary triple. The presentation is again adapted to our setting, and we refer to \cite{BGP} for general definitions.

As already mentioned, all self-adjoint extensions of $H_\alpha$ can be characterized by the $2\times 2$ matrices $C$ and $D$ satisfying  two simple conditions, and these extensions are denoted by $H_\alpha^{\CD}$. In the special case $(C,D) = (1,0)$, then $H_\alpha^{10}$ is equal to the original Aharonov-Bohm operator $H_\alpha^{\AB}$. Recall that this operator corresponds to the Friedrichs extension of $H_\alpha$ and that its spectrum is equal to $\R_+$. This operator is going to play a special role in the sequel.

Let us consider $\xi=(\xi_0,\xi_{-1})\in\CC^2$ and $z \in \CC \setminus \RR_+$. It is proved in \cite{BGP} that there exists a unique $\f\in\ker(H_\alpha^*-z)$ with $\Gamma_1 \f=\xi$. This solution is explicitly given by the formula: $\f:=\gamma(z)\xi$ with
\begin{equation*}
\gamma(z)\xi= \frac{\xi_0}{ N_{z,0}\,a_\alpha(z)}\, \f_{z,0}
+\frac{\xi_{-1}}{N_{z,-1}a_{1-\alpha}(z)} \f_{z,-1}
\end{equation*}
The Weyl function $M(z)$ is then defined by the relation
$M(z):=\Gamma_2 \;\!\gamma(z)$.
In view of the previous calculations one has
\begin{eqnarray*}
M(z) &=&  2 \left(\begin{smallmatrix}
\alpha\;  b_\alpha(z)/a_\alpha(z) & 0\\
0 & (1-\alpha)\; b_{1-\alpha}(z)/a_{1-\alpha}(z)
\end{smallmatrix}\right)\\
&=& - \frac{2}{\pi} \sin (\pi \alpha)\,
\left(\begin{smallmatrix}
\frac{\Gamma(1-\alpha)^2 e^{-i\pi\alpha}}{4^\alpha}(k^\alpha)^2 & 0 \\
0& \frac{\Gamma(\alpha)^2 e^{-i\pi(1-\alpha)}}{4^{1-\alpha}}(k^{1-\alpha})^2
\end{smallmatrix}\right).
\end{eqnarray*}
In particular, one observes that for $z \in \CC \setminus \RR_+$ one has $M(0):=\lim_{z\to 0} M(z)=0$.

In terms of the Weyl function and of the \emph{$\gamma$-field} $\gamma$ the Krein resolvent formula has the simple form:
\begin{multline}\label{Krein}
(H_\alpha^{\CD}-z)^{-1}-(H_\alpha^{\AB}-z)^{-1}=-\gamma(z)\big( D M(z)-C\big)^{-1} D \gamma(\Bar z)^*\\
=-\gamma(z)D^*\big(M(z) D^*-C^*\big)^{-1}\gamma(\Bar z)^*
\end{multline}
for $z \in \rho(H_\alpha^{\AB})\cap \rho(H_\alpha^{\CD})$. The following result is also derived within this formalism,
see \cite{AP} for i), \cite[Thm.~5]{DM} and the matrix reformulation \cite[Thm.~3]{GO} for ii).
In the statement, the equality $M(0)=0$ has already been taken into account.

\begin{lemma}
\begin{enumerate}
\item[i)] The value $z\in \R_-$ is an eigenvalue of $H_\alpha^{\CD}$ if and only if $\det \big(DM(z)-C\big) =0$, and in that case one has
\begin{equation*}
\ker(H_\alpha^{\CD}-z) = \gamma(z) \ker \big(DM(z)-C\big)\ .
\end{equation*}
\item[ii)] The number of negative eigenvalues of $H_\alpha^{\CD}$ coincides with the number of negative eigenvalues of the matrix $C D^*$.
\end{enumerate}
\end{lemma}

We stress that the number of eigenvalues does not depend on $\alpha \in (0,1)$, but only on the choice of $C$ and $D$.

Let us now add some comments about the role of the parameters $C$ and $D$ and discuss
some of their properties.
Two pairs of matrices $(C,D)$ and $(C',D')$ satisfying \eqref{eq-mcd} define the same boundary condition
({\it i.e.}~the same self-adjoint extension) if and only if there exists some invertible matrix
$L \in M_2(\CC)$ such that  $C'=LC$ and $D'=LD$ \cite[Prop.~3]{P}.
In particular, if $(C,D)$ satisfies \eqref{eq-mcd} and if $\det(D)\neq 0$,
then the pair $(D^{-1}C,1)$ defines the same boundary condition (and $D^{-1}C$
is self-adjoint).
Hence there is an arbitrariness in the choice of these parameters.
This can avoided in several ways.

First, one can establish a bijection between all boundary conditions
and the set $U(2)$ of the unitary $2\times 2$ matrices $U$ by setting
\begin{equation}
      \label{CDU}
C = C(U) := \frac{1}{2} (1-U) \quad \hbox{ and } \quad D = D(U) = \frac{i}{2}(1+U)\,,
\end{equation}
see a detailed discussion in \cite{Ha}. We shall comment more on this in the last section.

Another possibility is as follows ({\it cf.}~\cite{Pos} for details): There is a bijection between
the set of all boundary conditions and the set of triples $(\L,I,L)$,
where $\L\in \big\{\{0\},\CC,\CC^2\big\}$, $I:\L \to \CC^2$ is an identification map
(identification of $\L$ as a linear subspace of $\CC^2$) and $L$ is a self-adjoint operator in $\L$.
For example, given such a triple $(\L,I,L)$ the corresponding
boundary condition is obtained by setting
\begin{equation*}
C=C(\L,I,L):= L\oplus 1 \quad \hbox{and} \quad D=D(\L,I,L):=1\oplus 0
\end{equation*}
with respect to the decomposition $\CC^2=[I\L]\oplus [I\L]^\perp$.
On the other hand, for a pair $(C,D)$ satisfying \eqref{eq-mcd}, one can set
$\L:=\CC^d$ with $d := 2-\dim[\ker(D)]$,
$I: \L\to \CC^2$ is the identification map of $\L$ with $\ker(D)^\perp$ and $L:=(DI)^{-1} CI$.
In this framework, one can check by a direct calculation that for any $K \in M_2(\C)$ such that $DK-C$ is invertible,
one has
\begin{equation}
\label{eq-red0}
\big( D K-C\big)^{-1} D=I(PKI-L)^{-1}P,
\end{equation}
where $P:\C^2 \to \L$ is the adjoint of $I$, {\it i.e.}~the composition
of the orthogonal projection onto $I\L$ together with the identification of $I\L$ with $\L$.

Let us finally note that the conditions \eqref{eq-mcd} imply some specific properties related
to commutativity and adjointness. We shall need in particular:

\begin{lemma}\label{prop-minv}
Let $(C,D)$ satisfies \eqref{eq-mcd} and $K\in M_2(\CC)$ with $\Im K>0$. Then
\begin{enumerate}
\item[i)] The matrices $DK-C$ and $DK^*-C$ are invertible,
\item[ii)] The equality $\big[(DK-C)^{-1}D\big]^*=(DK^*-C)^{-1} D$ holds.
\end{enumerate}
\end{lemma}

\begin{proof}
i) By contraposition, let us assume that $\det(DK-C)=0$. Passing to the adjoint,
one also has $\det(K^*D^*-C^*)=0$, {\it i.e.}~there exists $f\in\CC^2$ such that
$K^*D^*f=C^*f$. By taking the scalar product with $D^*f$
one obtains that $\langle D^*f, K D^*f\rangle=\langle f,CD^*f\rangle$.
The right-hand side is real due to (i) in \eqref{eq-mcd}.
But since $\Im K>0$, the equality is possible if and only if $D^*f=0$.
It then follows that $C^*f=K^*D^*f=0$, which contradicts (ii) in \eqref{eq-mcd}.
The invertibility of $DK^*-C$ can be proved similarly.

ii) If $\det(D)\ne 0$, then the matrix $A:=D^{-1}C$ is self-adjoint and it follows that
\[
\big[(DK-C)^{-1}D\big]^*=\big[(K-A)^{-1}\big]^*=(K^*-A)^{-1}=(DK^*-C)^{-1}D\ .
\]
If $D=0$, then the equality is trivially satisfied.
Finally, if $\det (D)=0$ but $D\ne 0$ one has $\L:=\CC$. Furthermore, let us define
$I:\CC\to \CC^2$ by $I\L:=\ker(D)^\perp$ and let $P:\CC^2 \to \CC$ be its adjoint map.
Then, by the above construction there exists $\ell \in \R$ such that
$(DK-C)^{-1}D=I(PKI-\ell)^{-1}P$. It is also easily observed that
$PKI$ is just the multiplication by some $k\in \CC$ with $\Im k>0$, and hence
$(DK-C)^{-1}D=I(k-\ell)^{-1}P$.
Similarly one has $(DK^*-C)^{-1}D=I(\Bar{k}-\ell)^{-1}P$. Taking the adjoint
of the first expression leads directly to the expected equality.
\end{proof}

\section{Fourier transform and the dilation group}\label{secFour}

Before starting with the scattering theory, we recall some properties of the Fourier transform and of the dilation group in relation with the decomposition \eqref{decomposition}.
Let $\Fc$ be the usual Fourier transform, explicitly given on any $\f \in \HH$ and $y \in \R^2$ by
\begin{equation*}
[\Fc \f](y)= \frac{1}{2\pi}\;\!\li \int_{\R^2}\f(x)\;\!e^{-ix\cdot y}\;\!\d x
\end{equation*}
where $\li$~denotes the convergence in the mean.
Its inverse is denoted by $\Fc^*$. Since the Fourier transform maps the subspace $\HH_m$ of $\HH$ onto itself, we naturally set $\Fc_m: \HH_r \to \HH_r$ by the relation $\Fc(f\phi_m) = \Fc_m(f)\phi_m$ for any $f \in \HH_r$. More explicitly, the application $\Fc_m$ is the unitary map from $\HH_r$ to $\HH_r$ given on any $f \in \HH_r$ and almost every $\kappa \in \R_+$ by
\begin{equation*}
\hat{f}(\kappa):=[\Fc_m f](\kappa)= (-i)^{|m|}\;\!\li\int_{\R_+} r\;\!J_{|m|}(r\;\!\kappa)\;\!f(r)\;\!\d r\ ,
\end{equation*}
where $J_{|m|}$ denotes the Bessel function of the first kind and of order $|m|$. The inverse Fourier transform $\Fc_m^*$ is given by the same formula, with $(-i)^{|m|}$ replaced by $i^{|m|}$.

Now, let us recall that the unitary dilation group $\{U_\tau\}_{\tau \in \RR}$ is defined on any $\f \in \HH$ and $x \in \RR^2$ by
\begin{equation*}
[U_\tau \f](x) = e^\tau \f(e^\tau x)\ .
\end{equation*}
Its self-adjoint generator $A$ is formally given by $\frac{1}{2}(X\cdot (-i\nabla) + (-i\nabla)\cdot X)$, where $X$ is the position operator and $-i\nabla$ is its conjugate operator. All these operators are essentially self-adjoint on the Schwartz space on $\RR^2$.

An important property of the operator $A$ is that it leaves each subspace $\HH_m$ invariant. For simplicity, we shall keep the same notation for the restriction of $A$ to each subspace $\HH_m$.
So, for any $m \in \ZZ$, let $\varphi_m$ be an essentially bounded function on $\RR$. Assume furthermore that the family $\{\varphi_m\}_{m \in \ZZ}$ is bounded. Then the operator $\varphi(A):\HH \to \HH$ defined on $\HH_m$ by $\varphi_m(A)$ is a bounded operator in $\HH$.

Let us finally recall a general formula about the Mellin transform.
\begin{lemma}\label{Jensen}
Let $\varphi$ be an essentially bounded function on $\RR$ such that
its inverse Fourier transform is a distribution on $\RR$.
Then, for any $\f \in C^\infty_c\big(\RR^2\setminus\{0\}\big)$ one has
\begin{equation*}
[\varphi(A)\f](r,\theta) =
(2\pi)^{-1/2} \int_0^\infty\check{\varphi}
\big(-\ln(s/r)\big)\;\!\f(s,\theta)\;\!\frac{\d s}{r}\ ,
\end{equation*}
where the r.h.s.~has to be understood in the sense of distributions.
\end{lemma}

\begin{proof}
The proof is a simple application for $n=2$ of the general formulae developed
in \cite[p.~439]{Jen}.
Let us however mention that the convention of this reference on the minus sign
for the operator $A$ in its spectral representation
has not been adopted.
\end{proof}

As already mentioned $\varphi(A)$ leaves $\HH_m$ invariant.
More precisely, if $\f=f\phi_m$ for some $f \in C^\infty_c(\RR_+)$,
then $\varphi(A)\f = [\varphi(A)f]\phi_m$ with
\begin{equation}\label{formuleJen}
[\varphi(A)f](r) =
(2\pi)^{-1/2} \int_0^\infty\check{\varphi}
\big(-\ln(s/r)\big)\;\!f(s)\;\!\frac{\d s}{r}\ ,
\end{equation}
where the r.h.s.~has again to be understood in the sense of distributions

\section{Scattering theory}\label{secSca}

In this section we briefly recall the main definitions of the scattering theory,
and then give explicit formulae for the wave operators. The scattering operator
will be studied in the following section.

Let $H_1, H_2$ be two self-adjoint operators in $\HH$, and assume that the operator $H_1$ is purely absolutely continuous. Then the (time dependent) \emph{wave operators} are defined by the strong limits
\begin{equation*}
\Omega_\pm(H_2,H_1):=s-\lim_{t \to \pm \infty} e^{itH_2 } \;\!e^{-itH_1}
\end{equation*}
whenever these limits exist. In this case, these operators are isometries, and they are said \emph{complete} if their ranges are equal to the absolutely continuous subspace $\HH_{ac}(H_2)$ of $\HH$ with respect to $H_2$. In such a situation, the (time dependent)
\emph{scattering operator} for the system $(H_2,H_1)$ is defined by the product
\begin{equation*}
S(H_2,H_1):= \Omega_+^*(H_2,H_1)\;\!\Omega_-(H_2,H_1)
\end{equation*}
and is a unitary operator in $\HH$. Furthermore, it commutes with the operator $H_1$, and thus is unitarily equivalent to a family of unitary operators in the spectral representation of $H_1$.

We shall now prove that the wave operators exist for our model and that they are complete.
For that purpose, let us denote by $H_0:=-\Delta$ the Laplace operator on $\R^2$.

\begin{lemma}\label{existence}
For any self-adjoint extension $H_\alpha^{\CD}$, the wave operators
$\Omega_\pm(H_\alpha^{\CD},H_0)$
exist and are complete.
\end{lemma}

\begin{proof}
On the one hand, the existence and the completeness of the operators
$\Omega_\pm(H_\alpha^{\AB},H_0)$ has been proved in \cite{Rui}.
On another hand, the existence and the completeness of the operator
$\Omega_\pm(H_\alpha^{\CD},H_\alpha^{\AB})$ is well known since the difference of the resolvents
is a finite rank operator, see for example \cite[Sec.~X.4.4]{K}.
The statement of the lemma follows then by taking the chain rule \cite[Thm.~2.1.7]{Y}
and the Theorem 2.3.3 of  \cite{Y} on completeness into account.
\end{proof}

The derivation of the explicit formulae for the wave operators is based on the stationary approach, as presented in Sections 2.7 and 5.2 of \cite{Y}. For simplicity, we shall consider only $\Omega_-^{\CD}:=\Omega_-(H_\alpha^{\CD},H_0)$. For that purpose, let $\lambda \in \R_+$ and $\e>0$. We first study the expression
\begin{equation*}
\frac{\e}{\pi}\;\big\langle (H_0-\lambda + i\e)^{-1}\f,(H_\alpha^{\CD}-\lambda+ i\e)^{-1}\g \big\rangle
\end{equation*}
and its limit as $\e \to 0+$ for suitable $\f,\g \in \HH$ specified later on. By taking Krein resolvent formula into account, one can consider separately the two expressions:
\begin{equation*}
\frac{\e}{\pi}\;\big\langle (H_0-\lambda + i\e)^{-1}\f,(H_\alpha^{\AB}-\lambda+ i\e)^{-1}\g \big\rangle
\end{equation*}
and
\begin{equation*}
-\frac{\e}{\pi}\;\Big\langle (H_0-\lambda + i\e)^{-1}\f,
\gamma(\lambda- i\e)\big( D M(\lambda- i\e)-C\big)^{-1} D \gamma(\lambda+ i\e)^* \g \Big\rangle\ .
\end{equation*}
The first term will lead to the wave operator for the original Aharonov-Bohm system, as shown below.
So let us now concentrate on the second expression.

For simplicity, we set $z = \lambda + i\e$ and observe that
\begin{eqnarray*}
&&-\frac{\e}{\pi}\;\big\langle (H_0-\bar z)^{-1}\f,
\gamma(\bar z)\big( D M(\bar z)-C\big)^{-1} D \gamma(z)^* \g \big\rangle\\
=&&-\frac{\e}{\pi}\;\big\langle  \gamma(z)\big[\big( D M(\bar z)-C\big)^{-1} D\big]^*
\gamma(\bar z)^*(H_0-\bar z)^{-1}\f,\g \big\rangle\ .
\end{eqnarray*}
Then, for every $r \in \R_+$ and $\theta \in [0,2\pi)$ one has
\begin{eqnarray*}
&&-\frac{\e}{\pi}\Big[ \gamma(z)\big[\big( D M(\bar z)-C\big)^{-1} D\big]^*
\gamma(\bar z)^* (H_0-\bar z)^{-1} \f\Big](r,\theta) \\
=&&-\frac{\e}{\pi}
\left(\begin{smallmatrix}
H^{(1)}_\alpha(\sqrt{z}r)\phi_0(\theta) \\
H^{(1)}_{1-\alpha}(\sqrt{z}r)\phi_{-1}(\theta)
\end{smallmatrix}\right)^{\!\!T}
\cdot A(z) \big[\big( D M(\bar z)-C\big)^{-1} D\big]^* A(\bar z)^*
\left(\begin{smallmatrix}
\xi_0(z,\f) \\
\xi_{-1}(z,\f)
\end{smallmatrix}\right)
\end{eqnarray*}
with
\begin{equation*}
A(z) =
\begin{pmatrix}
a_\alpha(z)^{-1} & 0 \\
0 & a_{1-\alpha}(z)^{-1}
\end{pmatrix}
\end{equation*}
and
\begin{multline*}
\begin{pmatrix}
\xi_0(z,\f) \\
\xi_{-1}(z,\f)
\end{pmatrix}
=
\begin{pmatrix}
\big\langle
H^{(1)}_\alpha(\sqrt{\bar z}\cdot)\phi_0,(H_0-\bar z)^{-1}\f
\big\rangle \\
\big \langle
H^{(1)}_{1-\alpha}(\sqrt{\bar z}\cdot)\phi_{-1},(H_0-\bar z)^{-1} \f
\big \rangle
\end{pmatrix}
=
\begin{pmatrix}
\big\langle
\Fc (H_0- z)^{-1} H^{(1)}_\alpha(\sqrt{\bar z}\cdot)\phi_0,\hat{\f}
\big\rangle \\
\big \langle
\Fc (H_0- z)^{-1} H^{(1)}_{1-\alpha}(\sqrt{\bar z}\cdot)\phi_{-1},\hat{\f}
\big \rangle
\end{pmatrix}\\
=
\begin{pmatrix}
\big\langle
\Fc_0 (H_0- z)^{-1} H^{(1)}_\alpha(\sqrt{\bar z}\cdot),\hat{f}_0
\big\rangle_{\R_+} \\
\big \langle
\Fc_{-1} (H_0- z)^{-1} H^{(1)}_{1-\alpha}(\sqrt{\bar z}\cdot),\hat{f}_{-1}
\big \rangle_{\R_+}
\end{pmatrix}
=
\begin{pmatrix}
\big\langle
(X^2- z)^{-1}\Fc_0 H^{(1)}_\alpha(\sqrt{\bar z}\cdot),\hat{f}_0
\big\rangle_{\R_+} \\
\big \langle
(X^2- z)^{-1} \Fc_{-1} H^{(1)}_{1-\alpha}(\sqrt{\bar z}\cdot),\hat{f}_{-1}
\big \rangle_{\R_+}
\end{pmatrix}
\end{multline*}
where $\langle \cdot,\cdot\rangle_{\R_+}$ denotes the scalar product in $L^2(\R_+,r\;\!\d r)$.

We shall now calculate separately the limit as $\e \to 0$ of the different terms. We recall the convention that for $z \in \CC\setminus \RR_+$ on choose $k = \sqrt{z}$ with $\Im(z)>0$.
For $\lambda \in \R_+$ one sets $\lim_{\e \to 0+} \sqrt{\lambda +i\e} =: \kappa$ with $\kappa \in \R_+$.
We first observe that for $\nu \in (0,1)$ one has
\begin{equation*}
a_\nu(\lambda_+):=\lim_{\e \to 0+} a_\nu(\lambda +i\e) = -\frac{2^\nu i}{\sin (\pi\nu)\;\! \Gamma(1-\nu)}\;\!\kappa^{-\nu}
\end{equation*}
but
\begin{equation*}
a_\nu(\lambda_-):=\lim_{\e \to 0+} a_\nu(\lambda -i\e) = -\frac{2^\nu ie^{-i\pi\nu}}{\sin (\pi\nu)\;\! \Gamma(1-\nu)}\;\!\kappa^{-\nu} \ .
\end{equation*}
Similarly, one observes that
\begin{equation*}
M(\lambda_\pm):=\lim_{\e \to 0+}M(\lambda\pm i\e) = - \frac{2}{\pi} \sin (\pi \alpha)\,
\left(\begin{smallmatrix}
\frac{\Gamma(1-\alpha)^2 \;\!e^{\mp i\pi\alpha}}{4^\alpha}\kappa^{2\alpha} & 0 \\
0& \frac{\Gamma(\alpha)^2\;\! e^{\mp i\pi(1-\alpha)}}{4^{1-\alpha}}\kappa^{2(1-\alpha)}
\end{smallmatrix}\right).
\end{equation*}
Note that $M(\lambda_+)=M(\lambda_-)^*$.
Finally, the most elaborated limit is calculated in the next lemma.

\begin{lemma}\label{Dirac}
For $m \in \ZZ$, $\nu \in (0,1)$ and $f \in C^\infty_c(\R_+)$ one has
\begin{equation*}
\lim_{\e \to 0+}\e \big\langle
(X^2- z)^{-1}\Fc_m H^{(1)}_\nu(\sqrt{\bar z}\cdot),f
\big\rangle_{\R_+} = ie^{i\pi\nu/2}(-1)^{|m|}f(\kappa)\ .
\end{equation*}
\end{lemma}

\begin{proof}
Let us start by recalling that for $w \in \C$ satisfying $-\frac{\pi}{2}<\arg(w)\leq \pi$ one has \cite[eq.~9.6.4]{AS}~:
\begin{equation*}
H^{(1)}_\nu(w)=\frac{2}{i\pi}\;\!
e^{-i\pi\nu/2}\;\!K_\nu(-iw)\ ,
\end{equation*}
where $K_\nu$ is the modified Bessel function of the second kind and of order $\nu$.
Then, for $r \in \R_+$ it follows that (by using \cite[Sec.~13.45]{W} for the last equality)
\begin{eqnarray*}
&&[\Fc_m H^{(1)}_\nu(\sqrt{\bar z}\cdot) ](r) \\
=&&
(-i)^{|m|}\li \int_{\R_+}\rho J_{|m|}(r\rho)H^{(1)}_\nu\big(\sqrt{\bar z}\rho\big)\d \rho  \\
=&& (-i)^{|m|}\frac{2}{i\pi}\;\!
e^{-i\pi\nu/2}\li \int_{\R_+}\rho J_{|m|}(r\rho) K_\nu\big(-i\sqrt{\bar z}\rho\big)\d \rho \\
=&& (-i)^{|m|}\frac{2}{i\pi}\;\!
e^{-i\pi\nu/2}\frac{1}{r^2}\li \int_{\R_+}\rho J_{|m|}(\rho) K_\nu\Big(-i\frac{\sqrt{\bar z}}{r}\rho\Big)\d \rho \\
=&&c \frac{1}{r^2} \Big(-i\frac{\sqrt{\bar z}}{r}\Big)^{-2-|m|}
\F21\Big(\frac{|m|+\nu}{2}+1,
\frac{|m|-\nu}{2}+1;|m|+1;-\Big(-i\frac{\sqrt{\bar z}}{r}\Big)^{-2}\Big) \,
\end{eqnarray*}
where $\F21$ is the Gauss hypergeometric function \cite[Chap.~15]{AS} and
$c$ is given by
\begin{equation*}
c:=(-i)^{|m|}\frac{2}{i\pi}\;\!
e^{-i\pi\nu/2}\frac{\Gamma\big(\frac{|m|+\nu}{2}+1\big)\;\!\Gamma \big(\frac{|m|-\nu}{2}+1\big)}{\Gamma(|m|+1)}
\ .
\end{equation*}
Now, observe that $\Big(-i\frac{\sqrt{\bar z}}{r}\Big)^{-2-|m|} = -(-i)^{-|m|}
\Big(\frac{\sqrt{\bar z}}{r}\Big)^{-2-|m|}$ and
$-\Big(-i\frac{\sqrt{\bar z}}{r}\Big)^{-2} = \frac{r^2}{\bar z}$.
Thus, one has obtained
\begin{equation*}
[\Fc_m H^{(1)}_\nu(\sqrt{\bar z}\cdot) ](r) =
d\frac{1}{r^2} \Big(\frac{\sqrt{\bar z}}{r}\Big)^{-2-|m|}
\F21\Big(\frac{|m|+\nu}{2}+1,
\frac{|m|-\nu}{2}+1;|m|+1;\frac{r^2}{\bar z}\Big)
\end{equation*}
with
\begin{equation*}
d=-\frac{2}{i\pi}\;\!
e^{-i\pi\nu/2}\frac{\Gamma\big(\frac{|m|+\nu}{2}+1\big)\;\!\Gamma \big(\frac{|m|-\nu}{2}+1\big)}{\Gamma(|m|+1)}
\ .
\end{equation*}
By taking into account Equality 15.3.3 of \cite{AS} one can isolate from the $\F21$-function a factor which is singular when the variable goes to $1$:
\begin{eqnarray*}
&&\F21\Big(\frac{|m|+\nu}{2}+1,
\frac{|m|-\nu}{2}+1;|m|+1;\frac{r^2}{\bar z}\Big) \\
=&& \frac{1}{1-r^2\bar z^{-1}} \ \F21\Big(\frac{|m|+\nu}{2},
\frac{|m|-\nu}{2};|m|+1;\frac{r^2}{\bar z}\Big) \\
=&& -\frac{\bar z}{r^2-\bar z}\ \F21\Big(\frac{|m|+\nu}{2},
\frac{|m|-\nu}{2};|m|+1;\frac{r^2}{\bar z}\Big)\ .
\end{eqnarray*}
Altogether, one has thus obtained:
\begin{eqnarray*}
&&\e \big[
(X^2- z)^{-1}\Fc_m H^{(1)}_\nu(\sqrt{\bar z}\cdot)\big](r) \\
=&& -d \frac{\e}{(r^2-\bar z)(r^2-z)}
\frac{\bar z}{r^2} \Big(\frac{\sqrt{\bar z}}{r}\Big)^{-2-|m|}
\ \F21\Big(\frac{|m|+\nu}{2},
\frac{|m|-\nu}{2};|m|+1;\frac{r^2}{\bar z}\Big)\ .
\end{eqnarray*}

Now, observe that
\begin{equation*}
\frac{\e}{(r^2-\bar z)(r^2-z)} = \frac{\e}{(r^2-\lambda+i\e)(r^2-\lambda-i\e)} =
\frac{\e}{(r^2-\lambda)^2+\e^2}=:\pi\delta_\e(r^2-\lambda)
\end{equation*}
which converges to $\pi \delta(r^2-\lambda)$ in the sense of distributions on $\R$
as $\e$ goes to $0$.
Furthermore, the map
\begin{equation*}
\R_+\ni r \mapsto \F21\Big(\frac{|m|+\nu}{2},
\frac{|m|-\nu}{2};|m|+1;\frac{r^2}{\lambda -i\e}\Big)\in \CC
\end{equation*}
is locally uniformly convergent as $\e \to 0$ to a continuous function which is equal for $r=\kappa=\sqrt{\lambda}$ to $\Gamma(|m|+1)\big[\Gamma\big(\frac{|m|+\nu}{2}+1\big)\;\!
\Gamma \big(\frac{|m|-\nu}{2}+1\big)\big]^{-1}$.
By considering trivial extensions on $\R$, it follows that
\begin{eqnarray*}
&&\lim_{\e \to 0+}\e \big\langle
(X^2- z)^{-1}\Fc_m H^{(1)}_\nu(\sqrt{\bar z}\cdot),f
\big\rangle_{\R_+} \\
&=& -\bar d\pi\lim_{\e \to 0+} \int_{\R_+}r \delta_\e(r^2-\lambda)
 \overline{\frac{\bar z}{r^2} \Big(\frac{\sqrt{\bar z}}{r}\Big)^{-2-|m|}
\ \F21\Big(\frac{|m|+\nu}{2},
\frac{|m|-\nu}{2};|m|+1;\frac{r^2}{\bar z}\Big)} f(r)\d r\\
&=& -\frac{\bar d \pi}{2\kappa}
\kappa (-1)^{-|m|}
\frac{\Gamma(|m|+1)}{\Gamma\big(\frac{|m|+\nu}{2}+1\big)\;\!\Gamma \big(\frac{|m|-\nu}{2}+1\big)}
f(\kappa) \\
&=&ie^{i\pi\nu/2}(-1)^{|m|}f(\kappa)\ .
\end{eqnarray*}
\end{proof}

By adding these different results and by taking Lemma~\ref{prop-minv} into account,
one has thus proved:

\begin{lemma}\label{limite1}
For any $\f$ of the form $\sum_{m \in \ZZ}f_m \phi_m$ with  $f_m = 0$ except for a finite number of $m$ for which $\hat{f}_m \in C_c^\infty(\R_+)$ and for any $\g \in C_c^\infty(\R^2\setminus \{0\})$, one has
\begin{eqnarray*}
&&\lim_{\e \to 0+}\; -\frac{\e}{\pi}\;\Big\langle (H_0-\lambda + i\e)^{-1}\f,
\gamma(\lambda- i\e)\big( D M(\lambda- i\e)-C\big)^{-1} D \gamma(\lambda+ i\e)^* \g \Big\rangle \\
&=&-\frac{1}{\pi}\Big\langle
\Big(\begin{smallmatrix}
H^{(1)}_\alpha(\kappa \cdot)\phi_0 \\
H^{(1)}_{1-\alpha}(\kappa \cdot)\phi_{-1}
\end{smallmatrix}\Big)^{\!\!T}
\cdot A(\lambda_+) \big( D M(\lambda_+)-C\big)^{-1} D A(\lambda_-)^*
\left(\begin{smallmatrix}
ie^{i\pi\alpha/2}\hat{f}_0(\kappa) \\
-ie^{i\pi(1-\alpha)/2}\hat{f}_{-1}(\kappa)
\end{smallmatrix}\right),
\g\Big\rangle
\end{eqnarray*}
\end{lemma}

Before stating the main result on $\Omega_-^{\CD}$, let us first
present the explicit form of the stationary wave operator $\widetilde{\Omega}_-^{\AB}$.
Note that for this operator the equality between the time dependent approach and
the stationary approach is known \cite{AT,DS,Rui}, and that a preliminary version of the
following result has been given in \cite{R}.
So, let us observe that since the operator $H_\alpha^{\AB}$ leaves each subspace $\HH_m$
invariant \cite{Rui}, it gives rise to a sequence of channel operators
$H_{\alpha,m}^{\AB}$ acting on $\HH_m$.
The usual operator $H_0$ admitting a similar decomposition, the stationary wave operators
$\widetilde{\Omega}_\pm^{\AB}$ can be defined in each channel,
{\it i.e.}~separately for each $m \in \Z$.
Let us immediately observe that the angular part does not play any role
for defining such operators. Therefore, we shall omit it as long as it does
not lead to any confusion, and consider the channel wave operators
$\widetilde{\Omega}_{\pm,m}^{\AB}$ from $\HH_r$ to $\HH_r$.

The following notation will be useful: $\T:=\{z\in \CC \mid |z|=1\}$ and
\begin{equation*}
\delta_m^\alpha = \hbox{$\frac{1}{2}$}\pi\big(|m|-|m+\alpha|\big)
=\left\{\begin{array}{rl}
-\hbox{$\frac{1}{2}$}\pi\alpha & \hbox{if }\ m\geq 0 \\
\hbox{$\frac{1}{2}$}\pi\alpha & \hbox{if }\ m< 0
\end{array}\right.\ .
\end{equation*}

\begin{prop}\label{waveopAB}
For each $m \in \Z$, one has
\begin{equation*}
\Omega_{\pm,m}^{\AB}=\widetilde{\Omega}_{\pm,m}^{\AB} = \varphi_m^\pm(A)\ ,
\end{equation*}
with $\varphi_m^\pm \in C\big([-\infty,+\infty],\T\big)$ given explicitly by
\begin{equation}\label{niceGamma}
\varphi^\pm_m(x):=e^{\mp i\delta_m^\alpha}\;\!
\frac{\Gamma\big(\frac{1}{2}(|m|+1+ix)\big)}{\Gamma\big(\frac{1}{2}(|m|+1-ix)\big)}
\;\!
\frac{\Gamma\big(\frac{1}{2}(|m+\alpha|+1-ix)\big)}{\Gamma\big(\frac{1}{2}(|m+\alpha|+1+ix)\big)}
\end{equation}
 and satisfying $\varphi_m^\pm(\pm\infty)= 1$ and
$\varphi_m^\pm(\mp\infty) = e^{\mp 2i\delta_m^\alpha}$.
\end{prop}

\begin{proof}
As already mentioned, the first equality in proved in \cite{Rui}. Furthermore
it is also proved there that for any $f \in \HH_r$ and $r \in \RR_+$
one has
\begin{equation*}
[\Omega^{\AB}_{\pm,m}\;\! f](r)= i^{|m|}\;\li\int_{\RR_+} \kappa\;\!
J_{|m+\alpha|}(\kappa\;\!r)\;\!e^{\mp i\delta_m^\alpha}\;\![\Fc_m f](\kappa)\;\!\d
\kappa\ .
\end{equation*}
In particular, if $f \in C_c^\infty(\RR_+)$, this expression can be rewritten as
\begin{equation}
\label{fgeneral}
\begin{aligned}
&& s-\lim_{N\to \infty}e^{\mp i\delta_m^\alpha}\int_0^N \kappa \;\!J_{|m+\alpha|}(\kappa\;\!r)
\Big[\int_0^\infty s\;\! J_{|m|}(s\;\!\kappa)\;\!f(s)\;\!\d s\Big]\d \kappa \\
&=& s-\lim_{N\to \infty}e^{\mp i\delta_m^\alpha} \int_0^\infty s\;\!f(s)
\Big[\int_0^N \kappa\;\! J_{|m|}(s\;\!\kappa)\;\!J_{|m+\alpha|}(\kappa\;\!r)\;\!\d \kappa\Big] \d s \\
&=& s-\lim_{N\to \infty}e^{\mp i\delta_m^\alpha}\int_0^\infty \frac{s}{r}\;\!f(s)
\Big[\int_0^{Nr} \kappa\;\! J_{|m|}(\hbox{$\frac{s}{r}$}\;\!\kappa)\;\!J_{|m+\alpha|}(\kappa)\;\!\d \kappa\Big] \frac{\d s}{r} \\
&=& e^{\mp i\delta_m^\alpha} \int_0^\infty \frac{s}{r} \Big[
\int_0^\infty \kappa\;\! J_{|m|}(\hbox{$\frac{s}{r}$}\;\!\kappa)
\;\!J_{|m+\alpha|}(\kappa)\;\!\d \kappa\Big]f(s)\;\! \frac{\d s}{r}\ ,
\end{aligned}
\end{equation}
where the last term has to be understood in the sense of distributions on $\R_+$.
The distribution between square brackets has been computed in \cite[Prop.~2]{KR}
but we shall not use here its explicit form.

Now, by comparing \eqref{fgeneral} with \eqref{formuleJen}, one observes that the channel
wave operator $\Omega^{\AB}_{\pm,m}$ is equal on a dense set in $\HH_r$ to
$\varphi_m^\pm(A)$ for a function $\varphi_m^\pm$ whose inverse Fourier transform is
the distribution which satisfies for $y\in\RR$:
\begin{equation*}
\check{\varphi}_m^\pm (y) =
\sqrt{2\pi}\;\!e^{\mp i\delta_m^\alpha}\;\!e^{-y}\;\!\Big[
\int_0^\infty \kappa\;\! J_{|m|}(e^{-y}\;\!\kappa)
\;\!J_{|m+\alpha|}(\kappa)\;\!\d \kappa\Big]\ .
\end{equation*}
The Fourier transform of this distribution can be computed.
Explicitly one has (in the sense of distributions) :
\begin{eqnarray*}
\varphi_m^\pm(x)
&=& e^{\mp i\delta^\alpha_m}\;\!\int_\R e^{-ixy}\;\!
e^{-y}\Big [\int_{\R_+} \kappa\;\!J_{|m|}(e^{-y}\;\!\kappa)\;\!
J_{|m+\alpha|}(\kappa) \;\!\d \kappa \Big ] \d y \\
&=& e^{\mp i\delta^\alpha_m}\;\!
\int_{\R_+} \kappa^{(1-ix)-1}\;\!J_{|m+\alpha|}(\kappa)\;\!\d \kappa \;\!
\int_{\R_+} s^{(1+ix)-1}\;\! J_{|m|}(s)\;\!\d s
\end{eqnarray*}
which is the product of two Mellin transforms. The explicit form of these transforms are
presented in \cite[Eq.~10.1]{O} and a straightforward computation leads directly to the expression \eqref{niceGamma}.
The second equality of the statement follows then by a density argument.

The additional properties of $\varphi_m^\pm$ can easily be obtained by taking into account the equality $\Gamma(\overline{z})=\overline{\Gamma(z)}$ valid for any $z \in \CC$ as well as the asymptotic development of the function $\Gamma$ as presented in \cite[Eq.~6.1.39]{AS}.
\end{proof}

Since the wave operators $\Omega_{\pm}^{\AB}$ admit a decomposition into channel wave operators, so does the scattering operator. The channel scattering
operator $S_m^{\AB}:=(\Omega_{+,m}^{\AB})^*\;\Omega_{-,m}^{\AB}$, acting from $\HH_r$ to $\HH_r$, is simply given by
\begin{equation*}
S_m^{\AB} = \overline{\varphi_m^+}(A)\;\!\varphi_m^-(A)=e^{2i\delta_m^\alpha}\ .
\end{equation*}

Now, let us set $\HH_\2:=\HH_0\oplus\HH_{-1}$ which is clearly isomorphic to $\HH_r\otimes \C^2$, and consider the decomposition $\HH =\HH_\2 \oplus \HH_\2^\bot$. It follows from the considerations of Section \ref{secgeneral} that for any pair $(C,D)$ the operator $\widetilde{\Omega}_\pm^{\CD}$ is reduced by this decomposition and that $\widetilde{\Omega}_-^{\CD}\big|_{\HH_\2^\bot} = \Omega_-^{\CD}\big|_{\HH_\2^\bot} = \Omega_-^{\AB}\big|_{\HH_\2^\bot}$.
Since the form of $\Omega_-^{\AB}$ has been exposed above, we shall concentrate only of the restriction of $\widetilde{\Omega}_-^{\CD}$ to $\HH_\2$.
For that purpose, let us define a matrix valued function which is closely related to the scattering operator. For $\kappa \in \R_+$ we set
\begin{eqnarray} \label{tildeS}
\nonumber \widetilde{S}_\alpha^{\CD}(\kappa)
&:=& 2i\sin(\pi\alpha)
\left(\begin{matrix}
\frac{\Gamma(1-\alpha)\;\!e^{-i\pi\alpha/2}}{2^\alpha}\;\!\kappa^{\alpha} & 0 \\
0 & \frac{ \Gamma(\alpha)\;\!e^{-i\pi(1-\alpha)/2}}{2^{1-\alpha}}\;\!\kappa^{(1-\alpha)}
\end{matrix}\right)  \\
\nonumber &&\cdot \left( D\,
\left(\begin{matrix}
\frac{\Gamma(1-\alpha)^2 \;\!e^{ -i\pi\alpha}}{4^\alpha}\;\!\kappa^{2\alpha} & 0 \\
0& \frac{\Gamma(\alpha)^2\;\! e^{ -i\pi(1-\alpha)}}{4^{1-\alpha}}\;\!\kappa^{2(1-\alpha)}
\end{matrix}\right)
+\frac{\pi}{2\sin(\pi\alpha)}C\right)^{-1} D \\
&& \cdot
\left(\begin{matrix}
\frac{ \Gamma(1-\alpha)\;\!e^{-i\pi\alpha/2}}{2^\alpha}\;\!\kappa^{\alpha} & 0 \\
0 & -\frac{ \Gamma(\alpha)\;\!e^{-i\pi(1-\alpha)/2}}{2^{1-\alpha}}\;\!\kappa^{(1-\alpha)}
\end{matrix}\right)\ .
\end{eqnarray}

\begin{theorem}\label{formulewave}
For any pair $(C,D)$ satisfying \eqref{eq-mcd}, the restriction of the wave operator $\Omega_-^{\CD}$ to $\HH_\2$ satisfies the equality
\begin{equation}\label{yoyo}
\Omega_-^{\CD}\big|_{\HH_\2} = \widetilde{\Omega}_-^{\CD}\big|_{\HH_\2} = \Big(
\begin{smallmatrix}\varphi^-_0(A) & 0 \\ 0 & \varphi^-_{-1}(A) \end{smallmatrix}\Big) +
\Big(
\begin{smallmatrix}\tilde{\varphi}_0(A) & 0 \\ 0 & \tilde{\varphi}_{-1}(A) \end{smallmatrix}\Big)
\widetilde{S}^{\CD}_\alpha(\sqrt{H_0}),
\end{equation}
where $\tilde{\varphi}_m \in C\big([-\infty,+\infty],\C\big)$ for $m \in \{0,-1\}$. Explicitly, for every $x \in \R$, $\tilde{\varphi}_m(x)$ is given by
\begin{eqnarray*}
\frac{1}{2\pi}\;\!e^{-i\pi|m|/2} \;\!e^{\pi x/2}\;\!
\frac{\Gamma\big(\frac{1}{2}(|m|+1+ix)\big)}{\Gamma
\big(\frac{1}{2}(|m|+1-ix)\big)}  \Gamma\Big(\frac{1}{2}(1+|m+\alpha|-ix)\Big) \;\!\Gamma\Big(\frac{1}{2}(1-|m+\alpha|-ix)\Big)
\end{eqnarray*}
and satisfies $\tilde{\varphi}_m(-\infty)=0$ and $\tilde{\varphi}_m(+\infty)=1$.
\end{theorem}

\begin{proof}
a) The stationary representation $\widetilde{\Omega}_-^{\CD}$ is defined by the formula \cite[Def.~2.7.2]{Y}:
\begin{equation*}
\big\langle \widetilde{\Omega}_-^{\CD}\f,\g \big \rangle = \int_0^\infty\lim_{\e \to 0+}\frac{\e}{\pi}
\big\langle (H_0-\lambda + i\e)^{-1}\f,(H_\alpha^{\CD}-\lambda+ i\e)^{-1}\g \big\rangle\; \d \lambda
\end{equation*}
for any $\f$ of the form $\sum_{m \in \ZZ} f_m\phi_m$ with $f_m = 0$ except for a finite number of $m$ for which $\hat{f}_m \in C_c^\infty(\R_+)$ and $\g \in C_c^\infty(\R^2\setminus\{0\})$.
By taking Krein resolvent formula into account, we can first consider the expression
\begin{equation*}
\int_0^\infty \lim_{\e \to 0+} \frac{\e}{\pi}\big\langle (H_0-\lambda + i\e)^{-1}\f,(H_\alpha^{\AB}-\lambda+ i\e)^{-1}\g \big\rangle \; \d \lambda
\end{equation*}
which converges to \cite{AT,DS,Rui}:
\begin{equation*}
\int_0^\infty \Big \langle \sum_{m \in \ZZ} i^{|m|}\;\!e^{i\delta_m^\alpha} J_{|m+\alpha|}(\kappa\cdot)\;\!\hat{f}_m(\kappa)\;\!\phi_m ,\g
\Big\rangle \;\!\kappa\;\d \kappa\ .
\end{equation*}
This expression was the starting point for the formulae derived in Proposition \ref{waveopAB}. This leads to the first term in the r.h.s.~of \eqref{yoyo}.

b) The second term to analyze is
\begin{equation}\label{etape1}
-\int_0^\infty\lim_{\e \to 0+} \frac{\e}{\pi} \Big\langle (H_0-\lambda + i\e)^{-1}\f,
\gamma(\lambda- i\e)\big( D M(\lambda- i\e)-C\big)^{-1} D \gamma(\lambda+ i\e)^* \g \Big\rangle\; \d \lambda\ .
\end{equation}
By using then Lemma \ref{limite1} and by performing some simple calculations, one obtains that \eqref{etape1} is equal to
\begin{equation*}
 \int_0^\infty
\Big\langle
\Big(
\begin{smallmatrix}
\frac{1}{2}i^\alpha H^{(1)}_\alpha(\kappa \cdot)\phi_0 \\
\frac{1}{2}i^{1-\alpha} H^{(1)}_{1-\alpha}(\kappa \cdot)\phi_{-1}
\end{smallmatrix}
\Big)^{\!\!T}
\widetilde{S}_\alpha^{\CD}(\kappa)
\left(
\begin{smallmatrix}
\hat{f}_0(\kappa)\\
\hat{f}_{-1}(\kappa)
\end{smallmatrix}
\right),
\g\Big\rangle \kappa\; \d \kappa\ .
\end{equation*}

Now, it will be proved below that
the operator $T_m$ defined for $m \in \{0,-1\}$ on $\Fc^*[C^\infty_c(\R_+)]$ by
\begin{equation}\label{bientotfini}
[T_m f](r):=\frac{1}{2}i^{|m+\alpha|}\int_0^\infty H^{(1)}_{|m+\alpha|}(\kappa\;\!r) \;\![\Fc_m f](\kappa)\;\!\kappa \;\!\d \kappa
\end{equation}
satisfies the equality $T_m = \tilde{\varphi}_m(A)$ with $\tilde{\varphi}_m$ given in the above statement.
The stationary expression is then obtained by observing that
$\Fc^*\widetilde{S}_\alpha^{\CD}(\k)\Fc=
\widetilde{S}_\alpha^{\CD}(\sqrt{H_0})$, where $\widetilde{S}_\alpha^{\CD}(\k)$ is the operator of multiplication by the function $\widetilde{S}_\alpha^{\CD}(\cdot)$.
Finally, the equality between the time dependent wave operator and the stationary wave operator is a consequence of Lemma \ref{existence}
and of \cite[Thm.~5.2.4]{Y}.

c) By comparing \eqref{bientotfini} with \eqref{formuleJen}, one observes that the operator $T_m$ is equal on a dense set in $\HH_r$ to $\tilde{\varphi}_m(A)$ for a function $\tilde{\varphi}_m$ whose inverse Fourier transform is
the distribution which satisfies for $y\in\RR$:
\begin{equation*}
\check{\tilde{\varphi}}_m(y)= \frac{1}{2}\;\!\sqrt{2\pi}\;\!e^{-i\delta^\alpha_m}\;\!
e^{y}\;\!\int_{\R_+} \kappa\;\!H^{(1)}_{|m+\alpha|}(e^{y}\;\!\kappa)\;\!
J_{|m|}(\kappa)\;\!\d \kappa \ .
\end{equation*}
As before, the Fourier transform of this distribution can be computed.
Explicitly one has (in the sense of distributions) :
\begin{eqnarray*}
\tilde{\varphi}_m(x) &=& \frac{1}{2}\;\!e^{-i\delta^\alpha_m}\;\! \int_\R  \;\!e^{-ixy}\;\!e^y \Big[
\int_{\R_+}\kappa\;\!H^{(1)}_{|m+\alpha|}(e^{y}\;\!\kappa)\;\!
J_{|m|}(\kappa)\;\!\d \kappa\Big] \d y \\
&=& \frac{1}{2}\;\!e^{-i\delta^\alpha_m} \int_{\R_+}\kappa^{(1+ix)-1}\;\! J_{|m|}(\kappa)\;\!\d \kappa \;\!
\int_{\R_+} s^{(1-ix)-1}\;\!H^{(1)}_{|m+\alpha|}(s)\;\! \d s \;\! \\
&=& \frac{1}{2\pi}\;\!e^{-i\pi|m|/2} \;\!(-i)^{ix}\;\!
\frac{\Gamma\big(\frac{1}{2}(|m|+1+ix)\big)}{\Gamma\big(\frac{1}{2}(|m|+1-ix)\big)} \\
&& \cdot  \Gamma\big(\frac{1}{2}(1+|m+\alpha|-ix)\big)\;\!\Gamma\big(\frac{1}{2}(1-|m+\alpha|-ix)\big)\ .
\end{eqnarray*}
The last equality is obtained by taking into account the relation between the Hankel function $H^{(1)}_\nu$ and the Bessel function $K_\nu$ of the second kind as well as the Mellin transform of the functions $J_\nu$ and the function $K_\nu$ as presented in \cite[Eq.~10.1 \& 11.1]{O}.

The additional properties of $\tilde{\varphi}_m$ can easily be obtained by using the asymptotic development of the function $\Gamma$ as presented in \cite[Eq.~6.1.39]{AS}.
\end{proof}

\section{Scattering operator}\label{secScatOp}

In this section, we concentrate on the scattering operator and on its asymptotic
values for large and small energies.

\begin{prop}
The restriction of the scattering operator $S(H_\alpha^{\CD},H_0)$ to $\HH_\2$ is explicitly given by
\begin{equation*}
S(H_\alpha^{\CD},H_0)\big|_{\HH_\2}=
S_\alpha^{\CD}(\sqrt{H_0})
\quad\hbox{with}\quad
S_\alpha^{\CD}(\kappa):=\begin{pmatrix}
e^{-i\pi\alpha} & 0 \\
0 & e^{i\pi\alpha}
\end{pmatrix}
+ \widetilde{S}_\alpha^{\CD}(\kappa)\ .
\end{equation*}
\end{prop}

\begin{proof}
Let us first recall that the scattering operator can be obtained from $\Omega_-^{\CD}$ by the formula \cite[Prop.~4.2]{AJS}:
\begin{equation*}
s-\lim_{t\to+\infty}e^{itH_0}\;\!e^{-itH}\Omega_-^{\CD}=S(H_\alpha^{\CD},H_0).
\end{equation*}
We stress that the completeness has been taken into account for this equality.
Now, let us set $U(t):=e^{-it\ln(H_0)/2}$, where $\ln(H_0)$ is the self-adjoint operator obtained by functional calculus.
By the intertwining property of the wave operators and by the invariance principle, one also has
\begin{equation*}
s-\lim_{t\to+\infty}U(-t)\;\!\Omega_-^{\CD}\;\!U(t)=
S(H_\alpha^{\CD},H_0).
\end{equation*}

On the other hand, the operator $\ln(H_0)/2$ is the generator of translations in the spectrum of $A$, {\it i.e.}~$U(-t) \;\!\varphi(A)\;\!U(t) = \varphi(A+t)$ for any $\varphi:\RR \to \CC$. Since $\{U(t)\}_{t\in\R}$ is also reduced by the decomposition \eqref{decomposition}, it follows that
\begin{eqnarray*}
&&s-\lim_{t\to+\infty}U(-t)\;\Big[\Omega_-^{\CD}\big|_{\HH_\2}\Big]\;U(t) \\
=&& s-\lim_{t\to+\infty}U(-t)\;\!\Big[
\Big(
\begin{smallmatrix}\varphi^-_0(A) & 0 \\ 0 & \varphi^-_{-1}(A) \end{smallmatrix}\Big) +
\Big(
\begin{smallmatrix}\tilde{\varphi}_0(A) & 0 \\ 0 & \tilde{\varphi}_{-1}(A) \end{smallmatrix}\Big)
\widetilde{S}^{\CD}_\alpha(\sqrt{H_0})
\Big]\;\!U(t) \\
=&&
\Big(
\begin{smallmatrix}\varphi^-_0(+\infty) & 0 \\ 0 & \varphi^-_{-1}(+\infty) \end{smallmatrix}\Big) +
\Big(
\begin{smallmatrix}\tilde{\varphi}_0(+\infty) & 0 \\ 0 & \tilde{\varphi}_{-1}(+\infty) \end{smallmatrix}\Big)
\widetilde{S}^{\CD}_\alpha(\sqrt{H_0}).
\end{eqnarray*}
The initial statement is then obtained by taking the asymptotic values mentioned in Proposition \ref{waveopAB} and Theorem \ref{formulewave} into account.
\end{proof}

Even if the unitarity of the scattering operator follows from the general theory
we give below a direct verification in order to better understand its structure.
In the next statement, we only give the value of the scattering matrix at energy $0$
and energy equal to $+\infty$. However, more explicit expressions for $S^{\CD}_\alpha(\kappa)$
are exhibited in the proof.

\begin{prop}
The map
\begin{equation}\label{continu}
\RR_+\ni \kappa \mapsto S^{\CD}_\alpha(\kappa) \in M_2(\CC)
\end{equation}
is continuous, takes values in the set $U(2)$ and has explicit asymptotic values for $\kappa=0$ and $\kappa = +\infty$.
More explicitly, depending on $C, D$ or $\alpha$ one has:
\begin{enumerate}
\item[i)] If $D=0$, then $S^{\CD}_\alpha(\kappa)=\left(\begin{smallmatrix}
e^{-i\pi \alpha} & 0\\
0 & e^{i\pi\alpha}
\end{smallmatrix}\right)$,

\item[ii)] If $\det(D)\neq 0$, then
$S^{\CD}_\alpha(+\infty)=\left(\begin{smallmatrix}
e^{i\pi \alpha} & 0\\
0 & e^{-i\pi\alpha}
\end{smallmatrix}\right)$,

\item[iii)] If
$\dim[\ker(D)]=1$ and $\alpha =1/2$, then
$S^{\CD}_\alpha(+\infty)=(2\P -1)\;\!\left(\begin{smallmatrix}
i & 0\\
0 & -i
\end{smallmatrix}\right)$,
where $\P$ is the orthogonal projection onto $\ker(D)^\bot$,

\item[iv)] If $\ker(D)= \left(\begin{smallmatrix}
\CC\\ 0 \end{smallmatrix}\right)$ or if
$\dim[\ker(D)]=1$, $\alpha < 1/2$ and $\ker(D)\neq \left(\begin{smallmatrix}
0\\ \CC
\end{smallmatrix}\right)$,
then
$S^{\CD}_\alpha(+\infty)=\left(\begin{smallmatrix}
e^{-i\pi \alpha} & 0\\
0 & e^{-i\pi\alpha}
\end{smallmatrix}\right)$,

\item[v)] If $\ker(D)= \left(\begin{smallmatrix}
0\\ \CC
\end{smallmatrix}\right)$ or if
$\dim[\ker(D)]=1$, $\alpha > 1/2$ and $\ker(D)\neq \left(\begin{smallmatrix}
\CC\\ 0
\end{smallmatrix}\right)$,
then
$S^{\CD}_\alpha(+\infty)=\left(\begin{smallmatrix}
e^{i\pi \alpha} & 0\\
0 & e^{i\pi\alpha}
\end{smallmatrix}\right)$.
\end{enumerate}

Furthermore,
\begin{enumerate}

\item[a)] If  $C=0$, then
$S^{\CD}_\alpha(0)=\left(\begin{smallmatrix}
e^{i\pi \alpha} & 0\\
0 & e^{-i\pi\alpha}
\end{smallmatrix}\right)$,

\item[b)] If  $\det(C)\ne 0$, then
$S^{\CD}_\alpha(0)=\left(\begin{smallmatrix}
e^{-i\pi \alpha} & 0\\
0 & e^{i\pi\alpha}
\end{smallmatrix}\right)$,

\item[c)] If $\dim[\ker (C)]=1$ and $\alpha=1/2$, then
$S^{\CD}_\alpha(0)=
(1-2\Pi)\left(\begin{smallmatrix}
i & 0\\
0 & -i
\end{smallmatrix}\right)$,
where $\Pi$ is the orthogonal projection on $\ker(C)^\perp$.

\item[d)] If $\ker (C)=\left(\begin{smallmatrix}0\\ \CC \end{smallmatrix}\right)$ or if  $\dim[\ker(C)]=1$,
$\alpha>1/2$ and
$\ker (C)\ne \left(\begin{smallmatrix}\CC \\ 0 \end{smallmatrix}\right)$,
then
$S^{\CD}_\alpha(0)=\left(\begin{smallmatrix}
e^{-i\pi \alpha} & 0\\
0 & e^{-i\pi\alpha}
\end{smallmatrix}\right)$,

\item[e)] If $\ker (C)=\left(\begin{smallmatrix} \CC \\ 0\end{smallmatrix}\right)$ or if
$\dim[\ker (C)]=1$, $\alpha<1/2$ and $\ker (C)\ne \left(\begin{smallmatrix}0 \\ \CC \end{smallmatrix}\right)$,
then
$S^{\CD}_\alpha(0)=\left(\begin{smallmatrix}
e^{i\pi \alpha} & 0\\
0 & e^{i\pi\alpha}
\end{smallmatrix}\right)$.
\end{enumerate}
\end{prop}

\begin{proof}
Let us fix $\kappa>0$ and set $S:=S^{\CD}_\alpha(\kappa)$. For shortness, we also set $L:=\frac{\pi}{2\sin(\pi\alpha)}\;\!C$ and
\begin{gather*}
B=B(\kappa):=\left(\begin{smallmatrix}
\frac{\Gamma(1-\alpha)}{2^\alpha}\;\!\kappa^{\alpha} & 0 \\
0 & \frac{ \Gamma(\alpha)}{2^{1-\alpha}}\;\!\kappa^{(1-\alpha)}
\end{smallmatrix}\right),
\quad
\Phi:=\left(\begin{smallmatrix}
e^{-i\pi\alpha/2} & 0 \\
0 & e^{-i\pi(1-\alpha)/2}
\end{smallmatrix}\right),\quad
J:=\left(\begin{smallmatrix}
1 & 0 \\ 0 & -1
\end{smallmatrix}\right).
\end{gather*}
Note that the matrices $B$, $\Phi$ and $J$ commute with each other, that the matrix $B$
is self-adjoint and invertible, and that $J$ and $\Phi$ are unitary.

I) It is trivially checked that if $D=0$ the statement i) is satisfied.

II) Let us assume $\det(D)\neq 0$, {\it i.e.}~$D$ is invertible.
Without loss of generality and as explained at the end of Section \ref{boundaryt},
we assume than that $D=1$ and that the matrix $C$ is self-adjoint.
Then one has
\begin{eqnarray*}
S&=&\Phi^2\;\! J +2i\sin(\pi\alpha)\;\! B\;\! \Phi (B^2\;\!\Phi^2+L)^{-1} B\;\!\Phi\;\! J\\
&=&B \;\!\Phi\;\! (B^2\;\!\Phi^2+L)^{-1}\big[
B\;\!\big(\Phi^2+2i\sin(\pi\alpha)\big)+L\;\! B^{-1}\big]\Phi\;\!J.
\end{eqnarray*}
By taking the equality $\Phi^2+2i\sin(\pi\alpha)=\Phi^{-2}$ into account, it follows that
\begin{eqnarray*}
S&=& B\;\! \Phi\;\! (B^2\;\!\Phi^2+L)^{-1}\big(B\;\!\Phi^{-2}+L \;\!B^{-1}\big)\Phi \;\!J\\
&=& \Phi\;\!\big(\Phi^2+B^{-1}\;\! L\;\!B^{-1}\big)^{-1}
\big(\Phi^{-2}+B^{-1}\;\! L\;\!B^{-1}\big)\Phi \;\!J\\
&=& \Phi
\big(B^{-1}\;\! L\;\!B^{-1} +\cos(\pi\alpha)J -i\sin(\pi\alpha)\big)^{-1}
\big(B^{-1}\;\! L\;\!B^{-1} +\cos(\pi\alpha)J +i\sin(\pi\alpha)\big)
\Phi\;\! J \ .
\end{eqnarray*}
Since the matrix $B^{-1}\;\!L\;\!B^{-1} +\cos(\pi\alpha)J$ is self-adjoint, the above expression can be rewritten as
\begin{equation}\label{joliS}
S=\Phi \;\!\frac{B^{-1}\;\! L\;\!B^{-1} +\cos(\pi\alpha)J
+i\sin(\pi\alpha)}{B^{-1}\;\! L\;\!B^{-1} +\cos(\pi\alpha)J -i\sin(\pi\alpha)}
\;\!\Phi\;\!J
\end{equation}
which is clearly a unitary operator. The only dependence on  $\kappa$ in the terms $B$ is continuous and one has
\[
\lim_{\kappa \to +\infty}S^{\CD}_\alpha(\kappa) =\Phi \;\!\frac{\cos(\pi\alpha)J
+i\sin(\pi\alpha)}{\cos(\pi\alpha)J -i\sin(\pi\alpha)}
\;\!\Phi\;\!J = \left(\begin{smallmatrix}
e^{i\pi \alpha} & 0\\
0 & e^{-i\pi\alpha}
\end{smallmatrix}\right)
\]
which proves the statement ii)

III) We shall now consider the situation $\det(D)=0$ but $D\ne 0$.
Obviously, $\ker(D)$ is of dimension $1$. So let $p=(p_1,p_2)$ be a vector
in $\ker(D)$ with $\|p\|=1$. By \eqref{eq-red0} and by using the notation introduced
in that section one has
\begin{equation}\label{expressionS}
S=\Phi^2 \;\!J +2i\sin(\pi\alpha) \;\!B \;\!\Phi\;\! I\;\!(P\;\!B^2\;\!
\Phi^2\;\!I+\ell)^{-1} P\;\! B\;\!\Phi\;\! J.
\end{equation}
Note that the matrix of $\P:=I P:\CC^2 \to \CC^2$,
{\it i.e.}~the orthogonal projection
onto $p^\perp$,  is given by
\[
\P=\begin{pmatrix}
|p_2|^2 & - p_1 \Bar p_2\\
-\Bar p_1 p_2 & |p_1|^2
\end{pmatrix}
\]
and that $PB^2\Phi^2I$ is just the multiplication by the number
\begin{equation}\label{ck}
c(\kappa)=b_1^2(\kappa)\;\! |p_2|^2\;\!e^{-i\pi\alpha} - b_2^2(\kappa)\;\! |p_1|^2\;\!e^{i\pi\alpha},
\end{equation}
with  $b_1(\kappa)=\frac{\Gamma(1-\alpha)}{2^\alpha}\;\!\kappa^\alpha$ and
$b_2(\kappa)=\frac{ \Gamma(\alpha)}{2^{1-\alpha}}\;\!\kappa^{(1-\alpha)}$.

In the special case $\alpha=1/2$, the matrices $B$ and $\Phi$ have the special form
$B=\sqrt{\frac{\pi}{2}}\;\!\kappa^{1/2}$ and $\phi = e^{-i\pi/4}$.
Clearly, one also has $b_1 = b_2 = \sqrt{\frac{\pi}{2}}\;\!\kappa^{1/2}:=b$ and $c(\kappa)=-i\;\!b^2$.
In that case, the expression \eqref{expressionS} can be rewritten as
\begin{equation}
         \label{eq-a12}
S=i\left[\frac{\pi \;\!\kappa/2 -i\;\!\ell}{\pi \;\!\kappa/2 +i\;\!\ell} \;\!\P + (\P-1)
\right]\;\!J
\end{equation}
which is the product of unitary operators and thus is unitary. Furthermore,
the dependence in $\kappa$ is continuous and the asymptotic value is easily determined.
This proves statement iii)

If $\alpha \neq 1/2$, let us rewrite $S$ as
\begin{equation}\label{pourcorriger}
S= \Phi\;\!\big(c(\kappa)+\ell\big)^{-1}
\big[2\;\!i\;\!\sin(\pi\alpha)\;\!B\;\!\P\;\!B + c(\kappa)+\ell
\big] \;\!\Phi \;\!J \ .
\end{equation}
Furthermore, by setting $X_-:= \big(b_1^2\;\!|p_2|^2 - b_2^2\;\!|p_1|^2\big)$ and
$X_+:= \big(b_1^2\;\!|p_2|^2 + b_2^2\;\!|p_1|^2\big)$ one has
\[
c(\kappa)+\ell = \cos(\pi\alpha)\;\!X_- + \ell -i\sin(\pi\alpha)\;\!X_+
\]
and
\[
M:=
2\;\!i\;\!\sin(\pi\alpha)\;\!B\;\!\P\;\!B + c(\kappa)+\ell \;\! =
\begin{pmatrix}
e^{i\pi\alpha}\;\!X_- + \ell & -2\;\!i\;\!\sin(\pi\alpha)\;\!b_1\;\!b_2\;\!p_1\;\!\bar p_2 \\
-2\;\!i\;\!\sin(\pi\alpha)\;\!b_1\;\!b_2\;\!\bar p_1\;\! p_2 &
e^{-i\pi\alpha}\;\!X_- + \ell
\end{pmatrix}\ .
\]
With these notations, the unitary of $S$ easily follows from the equality $\det(M) = |c(\kappa)+\ell|^2$.
The continuity in $\kappa$ of all the expressions also implies the expected continuity of the map \eqref{continu}.
Finally, by taking \eqref{ck} and the explicit form of $M$ into account, the asymptotic
values of $S^{\CD}_\alpha(\kappa)$ for the cases iv) and v) can readily be obtained.

IV) Let us now consider the behavior of the scattering matrix near the zero energy.
If $C=0$, then $\det(D)\ne 0$ and one can use \eqref{joliS}
with $L=0$. The statement a) follows easily.

V) Assume that $\det(C)\ne 0$. In this case, it directly follows from \eqref{tildeS}
that $\widetilde S_\alpha^{\CD}(0)=0$, and then
$ S(0)=\left(\begin{smallmatrix}
e^{-i\pi\alpha} & 0\\
0 & e^{i\pi\alpha}
\end{smallmatrix}\right)$ which proves b).

VI) We now assume that $\dim[\ker(C)]=1$ and consider two cases.

Firstly, if $\det(D)\ne 0$ we can assume as in II) that $C$ is self-adjoint
and use again \eqref{joliS}. Introducing the entries of $L$,
\[
L=\begin{pmatrix}
l_{11} & l_{12}\\
\overline{l_{12}} & l_{22}
\end{pmatrix}
\]
one obtains
\begin{multline*}
\frac{B^{-1}\;\! L\;\!B^{-1} +\cos(\pi\alpha)J
+i\sin(\pi\alpha)}{B^{-1}\;\! L\;\!B^{-1} +\cos(\pi\alpha)J -i\sin(\pi\alpha)}=
\dfrac{1}{b_1^2 \;\!l_{22} \;\!e^{-i\pi\alpha} -b_2^2 \;\!l_{11}\;\! e^{i\pi\alpha}-b_1^2\;\! b_2^2}\\
\cdot\begin{pmatrix}
b_1^2\;\! l_{22}\;\! e^{i\pi\alpha} -b_2^2 \;\!l_{11}\;\! e^{i\pi\alpha}-b_1^2 \;\!b_2^2 \;\!e^{2i\pi\alpha}
& b_1\;\!b_2\;\! l_{12} \;\!(e^{-i\pi\alpha}-e^{i\pi\alpha})\\
b_1 \;\! b_2\;\! \overline{l_{12}} \;\!(e^{-i\pi\alpha}-e^{i\pi\alpha})
& b_1^2\;\! l_{22} \;\!e^{-i\pi\alpha} -b_2^2 \;\!l_{11}\;\! e^{-i\pi\alpha}-b_1^2 \;\!b_2^2 \;\!e^{-2i\pi\alpha}
\end{pmatrix}.
\end{multline*}
For $\alpha\ne 1/2$ one easily obtains the result stated in d) and e).
For $\alpha=1/2$, it follows that
\[
\lim_{\kappa\to 0+}\frac{B^{-1}\;\! L\;\!B^{-1} +\cos(\pi\alpha)J
+i\sin(\pi\alpha)}{B^{-1}\;\! L\;\!B^{-1} +\cos(\pi\alpha)J -i\sin(\pi\alpha)}=
\dfrac{2}{\tr (L)}\,L -1,
\]
and it only remains to observe that $L=\tr (L)\;\!\Pi$, where $\Pi$ is the orthogonal projection
on $\ker(L)^\perp=\ker(C)^\perp$. This proves c).

Secondly, let us assume that $\dim[\ker(D)]=1$. By \eqref{CDU} there exists
$U\in U(2)$ such that $\ker(C)=\ker(1-U)$ and $\ker(D)=\ker(1+U)$.
As a consequence, one has $\ker(C)=\ker(D)^\perp$ and then $\P=1-\Pi$.
On the other hand, we can use the expressions for the scattering operator
obtained in III). However, observe that $CI=C\big|_{\ker (D)^\perp}=C\big|_{\ker(C)}=0$
so we only have to consider these expressions in the special case $\ell=0$.
The asymptotic at $0$ energy are then easily deduced from these expressions.

By summing the results obtained for $\det(D)\neq 0$ and for $\dim[\ker(D)]=1$, and since
$D=0$ is not allowed if $\det(C)= 0$, one proves the cases c), d) and e).
\end{proof}

\begin{rem}
As can be seen from the proof,
the scattering matrix is independent of the energy in the following cases only:
\begin{itemize}
\item $D=0$, then $S_\alpha^{\CD}(\kappa)=\left(\begin{smallmatrix}
e^{-i\pi\alpha} & 0\\
0 & e^{i\pi\alpha}
\end{smallmatrix}\right)$,

\item $C=0$, then $S_\alpha^{\CD}(\kappa)=\left(\begin{smallmatrix}
e^{i\pi\alpha} & 0\\
0 & e^{-i\pi\alpha}
\end{smallmatrix}\right)$, see \eqref{joliS},

\item $\ker(C)=\ker(D)^\perp=\left(\begin{smallmatrix}\CC\\0\end{smallmatrix}\right)$,
then $S_\alpha^{\CD}(\kappa)=\left(\begin{smallmatrix}
e^{i\pi\alpha} & 0\\
0 & e^{i\pi\alpha}
\end{smallmatrix}\right)$, see \eqref{pourcorriger},

\item
$\ker(C)=\ker(D)^\perp=\left(\begin{smallmatrix}0\\ \CC \end{smallmatrix}\right)$,
then $S_\alpha^{\CD}(\kappa)=\left(\begin{smallmatrix}
e^{-i\pi\alpha} & 0\\
0 & e^{-i\pi\alpha}
\end{smallmatrix}\right)$, see \eqref{pourcorriger},

\item $\alpha=1/2$ and $\det(C)=\det(D)=0$,
then $S^{\CD}_\alpha(\kappa)=(2\P-1)\left(\begin{smallmatrix}
i & 0\\
0 & -i
\end{smallmatrix}\right)$,
where $\P$ is the orthogonal projection on $\ker(D)^\perp\equiv\ker(C)$,
see \eqref{eq-a12}.
\end{itemize}
\end{rem}

\section{Final remarks}\label{sec1to1}

As mentioned before, the parametrization of the self-adjoint extensions of $H_\alpha$ with the pair $(C,D)$ satisfying \eqref{eq-mcd} is highly none unique. For the sake of convenience, we recall here a one-to-one parametrization of all self-adjoint extensions and reinterpret a part of the results obtained before in this framework.

So, let $U \in U(2)$ and set
\begin{equation}\label{UtoCD}
C = C(U) := \frac{1}{2} (1-U) \quad \hbox{ and } \quad D = D(U) = \frac{i}{2}(1+U).
\end{equation}
It is easy to check that $C$ and $D$ satisfy both conditions \eqref{eq-mcd}.
In addition, two different elements $U,U'$ of $U(2)$ lead to two different self-adjoint operators $H_\alpha^{\CD}$ and $H_\alpha^{C'\!\!D'}$ with $C=C(U), D=D(U), C'=C(U')$ and $D'=D(U')$, {\it cf.}~\cite{Ha}.
Thus, without ambiguity we can write $H_\alpha^U$ for the operator $H_\alpha^{\CD}$ with $C,D$ given by \eqref{UtoCD}. Moreover, the set $\{H_\alpha^U\mid U \in U(2)\}$ describes all self-adjoint extensions of $H_\alpha$,
and, by \eqref{Krein}, the map $U\to H_\alpha^U$ is continuous in the norm resolvent topology.
Let us finally mention that the normalization of the above map has been chosen such that $H_\alpha^{-1}\equiv H_\alpha^{10}= H_\alpha^{\AB}$.

Obviously, we could use various parametrizations for the set $U(2)$. For example, one could set
\[
U = U(\eta,a,b)=e^{i\eta}\left(
\begin{matrix}
a & -\overline{b}\\
b & \overline{a}
\end{matrix}\right)
\]
with $\eta\in [0,2\pi)$ and $a,b \in \CC$ satisfying $|a|^2+|b|^2=1$, which is the parametrization used in \cite{AT}
(note nevertheless that the role of the unitary parameter was quite different).
We could also use the parametrization inspired by \cite{DS}:
\[
U = U(\omega,a,b,q)=e^{i\omega}\left(
\begin{matrix}
q\;\!e^{ia} & -(1-q^2)^{1/2}\;\!e^{-ib}\\
(1-q^2)^{1/2}\;\!e^{ib} & q\;\!e^{-ia}
\end{matrix}\right)
\]
with $\omega,a,b\in[0,2\pi)$ and $q \in [0,1]$. However, the following formulae look much simpler without such an arbitrary choice, and such a particularization can always be performed later on.

We can now rewrite part of the previous results in terms of $U$ :

\begin{lemma}
Let $U \in U(2)$. Then,
\begin{enumerate}
\item[i)] For $z \in \rho(H_\alpha^{\AB})\cap \rho(H_\alpha^U)$ the resolvent equation holds:
\begin{equation*}
(H_\alpha^U-z)^{-1}-(H_\alpha^{\AB}-z)^{-1}=-\gamma(z)\big[ (1+U)M(z)+i (1-U)\big]^{-1} (1+U) \gamma(\Bar z)^*\;,
\end{equation*}
\item[ii)] The number of negative eigenvalues of $H_\alpha^U$ coincides with the number of negative eigenvalues of the matrix
$i(U-U^*)$,
\item[iii)] The value $z\in \R_-$ is an eigenvalue of $H_\alpha^U$ if and only if $\det \big((1+U)M(z)+i (1-U)\big) =0$, and in that case one has
\begin{equation*}
\ker(H_\alpha^U-z) = \gamma(z) \ker \big((1+U)M(z)+i (1-U)\big)\; .
\end{equation*}
\end{enumerate}
\end{lemma}

The wave operators can also be rewritten in terms of the single parameter $U$.
We shall not do it here but simply express the asymptotic values of the scattering
operator $S^U_\alpha:=S(H_\alpha^U,H_0)$ in terms of $U$.
If $\lambda \in \C$ is an eigenvalue of $U$, we denote by $\V_\lambda$ the corresponding eigenspace.

\begin{prop}
One has:
\begin{enumerate}
\item[i)] If $U=-1$, then $S^U_\alpha(\kappa)\equiv S^{\AB}_\alpha=\left(\begin{smallmatrix}
e^{-i\pi \alpha} & 0\\
0 & e^{i\pi\alpha}
\end{smallmatrix}\right)$,

\item[ii)] If $-1 \not \in \sigma(U)$, then
$S^U_\alpha(+\infty)=\left(\begin{smallmatrix}
e^{i\pi \alpha} & 0\\
0 & e^{-i\pi\alpha}
\end{smallmatrix}\right)$,

\item[iii)] If $-1 \in \sigma(U)$ with multiplicity one
and $\alpha =1/2$, then
$S^U_\alpha(+\infty)=(2\P -1)\;\!\left(\begin{smallmatrix}
i & 0\\
0 & -i
\end{smallmatrix}\right)$,
where $\P$ is the orthogonal projection onto $\V_{-1}^\bot$,

\item[iv)] If
$\V_{-1}=\left(\begin{smallmatrix}
\CC\\ 0 \end{smallmatrix}\right)$ or if
 $-1 \in \sigma(U)$ with multiplicity one, $\alpha < 1/2$ and $\V_{-1}\neq \left(\begin{smallmatrix}
0\\ \CC
\end{smallmatrix}\right)$,
then
$S^U_\alpha(+\infty)=\left(\begin{smallmatrix}
e^{-i\pi \alpha} & 0\\
0 & e^{-i\pi\alpha}
\end{smallmatrix}\right)$,

\item[v)] If $\V_{-1}= \left(\begin{smallmatrix}
0\\ \CC
\end{smallmatrix}\right)$ or if
 $-1 \in \sigma(U)$ with multiplicity one, $\alpha > 1/2$ and $\V_{-1}\neq \left(\begin{smallmatrix}
\CC\\ 0
\end{smallmatrix}\right)$,
then
$S^U_\alpha(+\infty)=\left(\begin{smallmatrix}
e^{i\pi \alpha} & 0\\
0 & e^{i\pi\alpha}
\end{smallmatrix}\right)$.
\end{enumerate}

Furthermore,
\begin{enumerate}

\item[a)] If  $U=1$, then
$S^U_\alpha(0)=\left(\begin{smallmatrix}
e^{i\pi \alpha} & 0\\
0 & e^{-i\pi\alpha}
\end{smallmatrix}\right)$,

\item[b)] If  $1 \not \in \sigma(U)$, then
$S^U_\alpha(0)=\left(\begin{smallmatrix}
e^{-i\pi \alpha} & 0\\
0 & e^{i\pi\alpha}
\end{smallmatrix}\right)$,

\item[c)] If $1 \in \sigma(U)$ with multiplicity one and $\alpha=1/2$, then
$S^U_\alpha(0)=
(1-2\Pi)\left(\begin{smallmatrix}
i & 0\\
0 & -i
\end{smallmatrix}\right)$,
where $\Pi$ is the orthogonal projection on $\V_1^\perp$.

\item[d)] If $\V_1=\left(\begin{smallmatrix}0\\ \CC \end{smallmatrix}\right)$ or if
$1 \in \sigma(U)$ with multiplicity one, $\alpha>1/2$ and
$\V_1\ne \left(\begin{smallmatrix}\CC \\ 0 \end{smallmatrix}\right)$,
then
$S^U_\alpha(0)=\left(\begin{smallmatrix}
e^{-i\pi \alpha} & 0\\
0 & e^{-i\pi\alpha}
\end{smallmatrix}\right)$,

\item[e)] If $\V_1=\left(\begin{smallmatrix} \CC \\ 0\end{smallmatrix}\right)$ or if
$1 \in \sigma(U)$ with multiplicity one, $\alpha<1/2$ and
$\V_1\ne \left(\begin{smallmatrix}0 \\ \CC \end{smallmatrix}\right)$, then
$S^U_\alpha(0)=\left(\begin{smallmatrix}
e^{i\pi \alpha} & 0\\
0 & e^{i\pi\alpha}
\end{smallmatrix}\right)$.
\end{enumerate}
\end{prop}

\begin{rem}
The scattering matrix is independent of the energy in the following cases only:
\begin{itemize}
\item $U=-1$, then $S_\alpha^U(\kappa)\equiv S_\alpha^{\AB}=\left(\begin{smallmatrix}
e^{-i\pi\alpha} & 0\\
0 & e^{i\pi\alpha}
\end{smallmatrix}\right)$,

\item $U=1$, then $S_\alpha^U(\kappa)=\left(\begin{smallmatrix}
e^{i\pi\alpha} & 0\\
0 & e^{-i\pi\alpha}
\end{smallmatrix}\right)$, see \eqref{joliS},

\item $U=\left(\begin{smallmatrix}
1 & 0\\
0 & -1
\end{smallmatrix}\right)$,
then $S_\alpha^{U}(\kappa)=\left(\begin{smallmatrix}
e^{i\pi\alpha} & 0\\
0 & e^{i\pi\alpha}
\end{smallmatrix}\right)$, see \eqref{pourcorriger},

\item $U=\left(\begin{smallmatrix}
-1 & 0\\
0 & 1
\end{smallmatrix}\right)$,
then $S_\alpha^{U}(\kappa)=\left(\begin{smallmatrix}
e^{-i\pi\alpha} & 0\\
0 & e^{-i\pi\alpha}
\end{smallmatrix}\right)$, see \eqref{pourcorriger},

\item $\alpha=1/2$ and $\sigma(U)=\{-1,1\}$,
then $S^U_\alpha=(2\P-1)\left(\begin{smallmatrix}
i & 0\\
0 & -i
\end{smallmatrix}\right)$,
where $\P$ is the orthogonal projection on $\V_1$,
see \eqref{eq-a12}.
\end{itemize}
\end{rem}

\subsection*{Acknowledgment}
S. Richard is supported by the Swiss National Science Foundation.

\end{document}